\documentclass[superscriptaddress,prc,aps,twocolumn,showpacs]{revtex4}
\usepackage{amsmath,amssymb}
\usepackage{graphicx}
\newcommand{\Mainz}[1]
{\affiliation{Institut f\"ur Kernphysik, University of Mainz, D-55099 Mainz,Germany}}
\newcommand{\Bonn}[1]
{\affiliation{Helmholtz-Institut f\"ur Strahlen- und Kernphysik, University of Bonn,
 D-53115 Bonn, Germany}}
\newcommand{\Regina}[1]
{\affiliation{University of Regina, Regina, Saskatchewan S4S 0A2, Canada}}
\newcommand{\Glasgow}[1]
{\affiliation{SUPA School of Physics and Astronomy, University of Glasgow,
 Glasgow G12 8QQ, United Kingdom}}
\newcommand{\Kent}[1]
{\affiliation{Kent State University, Kent, Ohio 44242-0001, USA}}
\newcommand{\Giessen}[1]
{\affiliation{II Physikalisches Institut, University of Giessen, D-3539 Giessen, Germany}}
\newcommand{\Dubna}[1]
{\affiliation{Joint Institute for Nuclear Research, 141980 Dubna, Russia}}
\newcommand{\Pavia}[1]
{\affiliation{INFN Sezione di Pavia, I-27100 Pavia, Italy}}
\newcommand{\GWU}[1]
{\affiliation{The George Washington University, Washington, DC 20052-0001, USA}}
\newcommand{\LPI}[1]
{\affiliation{Lebedev Physical Institute, 119991 Moscow, Russia}}
\newcommand{\Dalhousie}[1]
{\affiliation{Dalhousie University, Halifax, Nova Scotia B3H 4R2, Canada}}
\newcommand{\Halifax}[1]
{\affiliation{Saint Mary’s University, Halifax, Nova Scotia B3H 3C3, Canada}}
\newcommand{\UniPavia}[1]
{\affiliation{Dipartimento di Fisica, Universit\`a di Pavia, I-27100 Pavia, Italy}}
\newcommand{\Basel}[1]
{\affiliation{Institut f\"ur Physik, University of Basel, CH-4056 Basel, Switzerland}}
\newcommand{\Edinburgh}[1]
{\affiliation{School of Physics, University of Edinburgh, Edinburgh EH9 3JZ,
 United Kingdom}}
\newcommand{\INR}[1]
{\affiliation{Institute for Nuclear Research, 125047 Moscow, Russia}}
\newcommand{\Sackville}[1]
{\affiliation{Mount Allison University, Sackville, New Brunswick E4L 1E6, Canada}}
\newcommand{\Zagreb}[1]
{\affiliation{Rudjer Boskovic Institute, HR-10000 Zagreb, Croatia}}
\newcommand{\Amherst}[1]
{\affiliation{University of Massachusetts, Amherst, Massachusetts 01003, USA}}
\newcommand{\UCLA}[1]
{\affiliation{University of California Los Angeles, Los Angeles, California 90095-1547, USA}}
\newcommand{\Jerusalem}[1]
{\affiliation{Racah Institute of Physics, Hebrew University of Jerusalem, Jerusalem 91904, Israel}}
\begin{document}
\title{
Measurement of the $\omega\to\pi^0e^+e^-$ and $\eta \to e^+e^-\gamma$ Dalitz
decays with the A2 setup at MAMI}

\author{P.~Adlarson}\Mainz \\
\author{F.~Afzal}\Bonn \\
\author{P.~Aguar-Bartolom\'e}\Mainz \\
\author{Z.~Ahmed}\Regina \\
\author{J.~R.~M.~Annand}\Glasgow \\
\author{H.~J.~Arends}\Mainz \\
\author{K.~Bantawa}\Kent \\
\author{R.~Beck}\Bonn \\
\author{H.~Bergh\"auser}\Giessen \\
\author{M.~Biroth}\Mainz \\
\author{N.~S.~Borisov}\Dubna \\
\author{A.~Braghieri}\Pavia \\
\author{W.~J.~Briscoe}\GWU \\
\author{S.~Cherepnya}\LPI \\
\author{F.~Cividini}\Mainz \\
\author{C.~Collicott}\Dalhousie \\ \Halifax \\
\author{S.~Costanza}\Pavia \\ \UniPavia \\
\author{I.~V.~Danilkin}\Mainz \\
\author{A.~Denig}\Mainz \\
\author{M.~Dieterle}\Basel \\
\author{E.~J.~Downie}\Mainz \\ \GWU \\
\author{P.~Drexler}\Mainz \\
\author{M.~I.~Ferretti Bondy}\Mainz \\
\author{L.~V.~Fil'kov}\LPI \\
\author{S.~Gardner}\Glasgow \\
\author{S.~Garni}\Basel \\
\author{D.~I.~Glazier}\Glasgow \\ \Edinburgh \\
\author{D.~Glowa}\Edinburgh \\
\author{W.~Gradl}\Mainz \\
\author{G.~M.~Gurevich}\INR \\
\author{D.~J.~Hamilton}\Glasgow \\
\author{D.~Hornidge}\Sackville \\
\author{G.~M.~Huber}\Regina \\
\author{T.~C.~Jude}\Edinburgh \\
\author{A.~K\"aser}\Basel\\
\author{V.~L.~Kashevarov}\Mainz \\ \LPI \\
\author{S.~Kay}\Edinburgh \\
\author{I.~Keshelashvili}\Basel\\
\author{R.~Kondratiev}\INR \\
\author{M.~Korolija}\Zagreb \\
\author{B.~Krusche}\Basel \\
\author{A.~Lazarev}\Dubna \\
\author{J.~Linturi}\Mainz \\
\author{V.~Lisin}\LPI \\
\author{K.~Livingston}\Glasgow \\
\author{I.~J.~D.~MacGregor}\Glasgow \\
\author{R.~Macrae}\Glasgow \\
\author{D.~M.~Manley}\Kent \\ 
\author{P.~P.~Martel}\Mainz \\ \Amherst \\
\author{J.~C.~McGeorge}\Glasgow \\
\author{E.~F.~McNicoll}\Glasgow \\
\author{V.~Metag}\Giessen \\
\author{D.~G.~Middleton}\Mainz \\ \Sackville \\
\author{R.~Miskimen}\Amherst \\
\author{E.~Mornacchi}\Mainz \\
\author{C.~Mullen}\Glasgow \\
\author{A.~Mushkarenkov}\Pavia \\ \Amherst \\ 
\author{A.~Neganov}\Dubna \\
\author{A.~Neiser}\Mainz \\ 
\author{A.~Nikolaev}\Bonn \\ 
\author{M.~Oberle}\Basel \\
\author{M.~Ostrick}\Mainz \\  
\author{P.~Ott}\Mainz \\   
\author{P.~B.~Otte}\Mainz \\
\author{D.~Paudyal}\Regina \\
\author{P.~Pedroni}\Pavia \\
\author{A.~Polonski}\INR \\  
\author{S.~Prakhov}\thanks{corresponding author, e-mail: prakhov@ucla.edu}\Mainz \\ \UCLA \\
\author{A.~Rajabi}\Amherst \\
\author{J.~Robinson}\Glasgow \\
\author{G.~Ron}\Jerusalem \\
\author{G.~Rosner}\Glasgow \\
\author{T.~Rostomyan}\Basel \\
\author{C.~Sfienti}\Mainz \\
\author{M.~H.~Sikora}\Edinburgh \\
\author{V.~Sokhoyan}\Mainz \\ \GWU \\
\author{K.~Spieker}\Bonn \\
\author{O.~Steffen}\Mainz \\
\author{I.~I.~Strakovsky}\GWU \\
\author{B.~Strandberg}\Glasgow \\
\author{Th.~Strub}\Basel \\
\author{I.~Supek}\Zagreb \\
\author{A.~Thiel}\Bonn \\
\author{M.~Thiel}\Mainz \\
\author{A.~Thomas}\Mainz \\   
\author{M.~Unverzagt}\Mainz \\ 
\author{Yu.~A.~Usov}\Dubna \\
\author{S.~Wagner}\Mainz \\
\author{N.~Walford}\Basel \\ 
\author{D.~P.~Watts}\Edinburgh \\
\author{D.~Werthm\"uller}\Glasgow \\ \Basel \\
\author{J.~Wettig}\Mainz \\
\author{L.~Witthauer}\Basel \\ 
\author{M.~Wolfes}\Mainz \\
\author{L.~A.~Zana}\Edinburgh \\

\collaboration{A2 Collaboration at MAMI}

\date{\today}
         
\begin{abstract}
 The Dalitz decays $\eta \to e^+e^-\gamma$ and $\omega\to \pi^0e^+e^-$ have
 been measured in the $\gamma p\to \eta p$ and $\gamma p\to \omega p$ reactions,
 respectively, with the A2 tagged-photon facility at the Mainz Microtron, MAMI.
 The value obtained for the slope parameter
 of the electromagnetic transition form factor of $\eta$,
 $\Lambda^{-2}_{\eta}=(1.97\pm 0.11_{\mathrm{tot}})$~GeV$^{-2}$,
 is in good agreement with previous measurements of the
 $\eta\to e^+e^-\gamma$ and $\eta \to \mu^+\mu^-\gamma$ decays.
 The uncertainty obtained in the value of $\Lambda^{-2}_{\eta}$
 is lower than in previous results based on the $\eta\to e^+e^-\gamma$ decay.
 The value obtained for the $\omega$ slope parameter,
 $\Lambda^{-2}_{\omega\pi^0}=(1.99\pm 0.21_{\mathrm{tot}})$~GeV$^{-2}$,
 is somewhat lower than previous measurements based on
 $\omega\to\pi^0\mu^+\mu^-$, but the results for the $\omega$
 transition form factor are in better agreement with theoretical
 calculations, compared to earlier experiments.
\end{abstract}

\pacs{
 14.40.Be, 
 13.20.-v, 
 13.40.Gp  
}

\maketitle

\section{Introduction}

 The electromagnetic (e/m) transition form factors (TFFs) of light
 mesons play an important role in understanding the properties
 of these particles as well as in low-energy precision
 tests of the Standard Model (SM) and Quantum Chromodynamics (QCD)~\cite{TFFW12}.
 In particular, these TFFs enter as contributions to the hadronic
 light-by-light (HLbL) scattering calculations~\cite{Colangelo_2014,Colangelo_2015}
 that are important for more accurate theoretical
 determinations of the anomalous magnetic moment of the muon, $(g-2)_\mu$,
 within the SM~\cite{g_2,Nyffeler_2016}.
 Recently, data-driven approaches, using dispersion relations, have been
 proposed~\cite{Colangelo_2014,Colangelo_2015,Pauk_2014} to make a substantial and
 model-independent improvement to the determination of the HLbL contribution to $(g-2)_\mu$.
 The precision of the calculations used to describe
 the HLbL contributions to $(g-2)_\mu$ can then be tested by directly comparing
 theoretical predictions from these approaches for e/m TFFs
 of light mesons with experimental data.
 The precise knowledge of TFFs for light mesons is essential for precision
 calculation of the decay rates of those mesons
 in rare dilepton modes, $e^+e^-$ and $\mu^+\mu^-$~\cite{Leupold_2015,Pere_1512}.
 So far there are discrepancies between theoretical calculations
 and experimental data for these rare decays, and the Dalitz decays
 of the corresponding mesons in the timelike (the energy transfer larger than
 the momentum transfer) momentum-transfer ($q$) region
 can be used in such calculations for both the normalization of these rare decays
 and as a background. The same applies to rare decays of light mesons into four leptons. 

\subsection{Amplitudes for Dalitz decays}
\label{Intro-DalitzDecays}
 For a structureless (pointlike) meson $A$, its decays into a lepton pair plus
 a photon or another meson, $A\to \ell^+\ell^-B$, can be described within
 Quantum Electrodynamics (QED) via $A\to \gamma^*B$, with the virtual photon $\gamma^*$
 decaying into the lepton pair~\cite{QED}. QED predicts a specific strong dependence
 of the meson-$A$ decay rate on the dilepton invariant mass, $m_{\ell\ell}=q$.
 A deviation from the pure QED dependence, caused by the actual electromagnetic
 structure of the meson $A$, is formally described by its e/m TFF~\cite{Landsberg}.
 The Vector-Meson-Dominance (VMD) model~\cite{Sakurai} can be used to describe the coupling of
 the virtual photon $\gamma^*$ to the meson $A$ via an intermediate virtual vector meson $V$.
 This mechanism is especially strong in the timelike momentum-transfer region,
 $(2m_{\ell})^2 < q^2 < m_A^2$, where a resonant behavior near
 $q = m_V$ of the virtual photon arises because the virtual vector meson is approaching
 the mass shell~\cite{Landsberg}, or even reaching it, as for the $\eta'\to \ell^+\ell^-\gamma$ decay.    
 Thus, measuring TFFs of light mesons is ideally suited for testing the VMD model.

 Experimentally, timelike TFFs can be determined by measuring the actual decay rate
 of $A\to \ell^+\ell^-B$ as a function of the dilepton invariant mass $m_{\ell\ell}=q$,
 normalizing this dependence to the partial decay width $\Gamma(A\to B\gamma)$,
 and then taking the ratio to the pure QED dependence for the decay rate of
 $A\to \gamma^*B \to \ell^+\ell^-B$.
 Based on QED, the decay rate of $\eta\to \gamma^*\gamma \to \ell^+\ell^-\gamma$ can
 be parametrized as~\cite{Landsberg} 
\begin{eqnarray}
 & & \frac{d\Gamma(\eta\to \ell^+\ell^-\gamma)}{dm_{\ell\ell}\Gamma(\eta\to \gamma\gamma)} =
 \frac{4\alpha}{3\pi m_{\ell\ell}} \times
\nonumber 
\\ 
 & \times &  (1-\frac{4m^2_{\ell}}{m^2_{\ell\ell}})^{\frac{1}{2}}
  (1+\frac{2m^2_{\ell}}{m^2_{\ell\ell}}) (1-\frac{m^2_{\ell\ell}}{m^2_{\eta}})^{3}
  |F_{\eta}(m_{\ell\ell})|^2 =
\nonumber 
\\ 
 & = & [{\rm{QED}}_{\eta}]  |F_{\eta}(m_{\ell\ell})|^2,
\label{eqn:dgdm_eta}
\end{eqnarray}
 where $F_{\eta}$ is the TFF of the $\eta$ meson and $m_{\eta}$ is the mass
 of the $\eta$ meson.

 Another feature of the $A\to \gamma^*B \to \ell^+\ell^-B$ decay amplitude is
 an angular anisotropy of the virtual photon decaying into a lepton pair,
 which also determines the density of events along $m^2(B \ell)$
 of the $A\to \ell^+\ell^-B$ Dalitz plot. For the $\ell^+$, $\ell^-$, and $B$
 in the rest frame of $A$, the angle $\theta^*$ between the direction of one
 of the leptons in the virtual-photon (or the dilepton) rest frame and the direction
 of the dilepton system (which is opposite to the direction of $B$)
 follows the dependence~\cite{NA60_2016}:
\begin{equation}
 f(\cos\theta^*) = 1 + \cos^2\theta^* + (\frac{2m_{\ell}}{m_{\ell\ell}})^2 \sin^2\theta^*,
\label{eqn:dtheta}
\end{equation}
 with the $\sin^2\theta^*$ term becoming very small when $m_{\ell\ell}>>2m_{\ell}$.

 The decay rate of $\omega\to \pi^0\gamma^* \to \pi^0\ell^+\ell^-$ can
 be parametrized as~\cite{Landsberg}
\begin{eqnarray}
 & & \frac{d\Gamma(\omega\to \pi^0\ell^+\ell^-)}{dm_{\ell\ell}\Gamma(\omega\to \pi^0\gamma)} =
 \frac{2\alpha}{3\pi m_{\ell\ell}}
 (1-\frac{4m^2_{\ell}}{m^2_{\ell\ell}})^{\frac{1}{2}} (1+\frac{2m^2_{\ell}}{m^2_{\ell\ell}}) \times
\nonumber 
\\ 
 & \times & [(1+\frac{m^2_{\ell\ell}}{m^2_{\omega}-m^2_{\pi^0}})^2
   -\frac{4m^2_{\omega}m^2_{\ell\ell}}{(m^2_{\omega}-m^2_{\pi^0})^2}]^{3/2}
  |F_{\omega\pi^0}(m_{\ell\ell})|^2 =
\nonumber 
\\ 
 & = & [{\rm{QED}}_{\omega\pi^0}]  |F_{\omega\pi^0}(m_{\ell\ell})|^2,
\label{eqn:dgdm_omega}
\end{eqnarray}
 where $F_{\omega\pi^0}$ is the $\omega\to \pi^0\gamma^*$ TFF, and $m_{\omega}$
 and $m_{\pi^0}$ are the masses of the $\omega$ and $\pi^0$ mesons, respectively.
 The angular dependence of the virtual photon decaying into a lepton pair
 for $\omega\to \pi^0\gamma^* \to \pi^0\ell^+\ell^-$ is the same as Eq.~(\ref{eqn:dtheta}).
 
 Note that the $[{\rm{QED}}(m_{\ell\ell})]$ terms in
 Eqs.~(\ref{eqn:dgdm_eta}) and (\ref{eqn:dgdm_omega}) and
 the angular dependence in Eq.~(\ref{eqn:dtheta}) represent only
 the leading-order term of the decay amplitudes, and, in principle,
 radiative corrections need to be considered for a more accurate calculation
 of $[{\rm{QED}}(m_{\ell\ell},\cos\theta^*)]$. Taking those corrections
 into account is vital for measuring the Dalitz decay
 $\pi^0\to e^+e^-\gamma$, where the magnitude of the corrections
 at the largest $q$ is even larger than the expected TFF contribution. 
 The most recent calculations of radiative corrections to the differential
 decay rate of the Dalitz decay $\pi^0\to e^+e^-\gamma$ were reported by
 the Prague group in Ref.~\cite{Husek_2015}. The authors of that work also mentioned
 that radiative corrections for $\eta\to e^+e^-\gamma$ could be evaluated
 by replacing the $\pi^0$ mass with the $\eta$ mass in their code. More precise
 calculations for $\eta\to e^+e^-\gamma$ by the Prague group are still
 in progress. Typically, taking radiative corrections into account
 makes the angular dependence of the virtual-photon decay weaker.    
 The corrected $[{\rm{QED}}_{\eta}]$ term integrated over $\cos\theta^*$ is $\sim$1.5\%
 larger than the leading-order term at low $q$ and becomes $\sim$10\% lower
 at $q=455$~MeV. The magnitude of radiative corrections for $[{\rm{QED}}_{\omega\pi^0}]$
 is expected to be of the same order.

 From the VMD assumption, TFFs are usually parametrized in a pole 
 approximation 
\begin{equation}
 F(m_{\ell\ell}) = (1-\frac{m^2_{\ell\ell}}{\Lambda^2})^{-1},
\label{eqn:Fm}
\end{equation}
 where $\Lambda$ is the effective mass of the virtual vector meson,
 and the parameter $\Lambda^{-2}$ reflects the TFF slope at $m_{\ell\ell}=0$.
 A simple VMD model would incorporate only the $\rho$, $\omega$, and $\phi$
 resonances (in the narrow-width approximation) as virtual vector mesons
 driving the photon interaction in $A\to \gamma^*B$. Using a quark model
 for the corresponding couplings would yield the TFF slope
 $\Lambda^{-2}_{\eta} = 1.80$~GeV$^{-2}$ and
 $\Lambda^{-2}_{\omega\pi^0} = 1.68$~GeV$^{-2}$~\cite{Landsberg},
 corresponding to $\Lambda_{\eta} = 745$~MeV and $\Lambda_{\omega\pi^0} = 772$~MeV.
 The nearness of $\Lambda_{\omega\pi^0}$ to the $\rho$ mass comes from
 isospin conservation in the $\omega \to \pi^0 \gamma^* \to \pi^0 \ell^+\ell^-$ decay,
 allowing only $\gamma^*\to \ell^+\ell^-$ with $I=1$, which eliminates contributions
 from $\omega$ and $\phi$ with $I=0$.

\subsection{Dalitz decays of $\eta$}
\label{Intro-EtaDecay}
 From the experimental and phenomenological point of view,
 the $\eta \to \gamma^* \gamma \to \ell^+\ell^-\gamma$ TFF is currently the one
 investigated most. The early measurement of the $\eta$ slope parameter by
 Lepton-G~\cite{Lepton_G_eta}, $\Lambda^{-2}_{\eta}=(1.90\pm0.40_{\rm{tot}})$~GeV$^{-2}$,
 was based on quite limited statistics.
 The first results from the NA60 Collaboration~\cite{NA60_2009},
 $\Lambda^{-2}_{\eta}=(1.95\pm 0.17_{\mathrm{stat}}\pm 0.05_{\mathrm{syst}})$~GeV$^{-2}$,
 was based on $2.6\cdot 10^4$ $\mu^+\mu^-$ pairs detected in peripheral
 In--In data, $9\cdot 10^3$ of which were identified
 to be from $\eta \to \mu^+\mu^-\gamma$ decays.
 The latest experiment by the NA60 Collaboration with p--A collisions~\cite{NA60_2016},
 which increased the statistics of muon pairs by one order of magnitute, reported
 $\Lambda^{-2}_{\eta}=(1.934\pm 0.067_{\mathrm{stat}}\pm 0.050_{\mathrm{syst}})$~GeV$^{-2}$,
 improving significantly the accuracy, compared to the earlier result. 
 The first measurement by the A2 Collaboration at MAMI,  
 $\Lambda^{-2}_{\eta}=(1.92\pm 0.35_{\mathrm{stat}}\pm 0.13_{\mathrm{syst}})$~GeV$^{-2}$,
 was based on an analysis of
 $1.35\cdot 10^3$ $\eta \to e^+e^-\gamma$ decays~\cite{eta_tff_a2_2011}.
 Later on, a higher-accuracy result,
 $\Lambda^{-2}_{\eta}=(1.95\pm 0.15_{\mathrm{stat}}\pm 0.10_{\mathrm{syst}})$~GeV$^{-2}$,
 obtained by the A2 Collaboration,
 was based on an analysis of $2.2\cdot 10^4$
 $\eta \to e^+e^-\gamma$ decays from a total of $3\cdot 10^7$ $\eta$
 mesons produced in the $\gamma p\to \eta p$ reaction~\cite{eta_tff_a2_2014}.
 In that work, there is also a detailed discussion of agreement between the experimental
 data and recent calculations available for the $\eta$ TFF at the moment.
 Combining those A2 results with available experimental data in the spacelike
 (the energy transfer less than the momentum transfer) region allowed the Mainz
 theoretical group to extract the slope parameter with the smallest uncertainty,
 $\Lambda^{-2}_{\eta} = (1.919 \pm 0.039)$~GeV$^-2$~\cite{Escribano_2015}.
 Such synergy between theory and experiment allowed
 a data-driven calculation of the $\eta \to \ell^+\ell^-$ rare
 decay~\cite{Pere_1512} and the reduction of the uncertainty in
 the pseudoscalar-exchange HLbL contribution
 to $(g-2)_\mu$~\cite{Sanchez-Puertas_2015}.
 The most recent $\eta \to \gamma^* \gamma$ calculation with
 the updated dispersive analysis by the J\"ulich group was presented
 in Ref.~\cite{Xiao_2015}, demonstrating even better agreement with the data,
 compared to the previous calculations by this group in Ref.~\cite{Hanhart}.
 The improvement was based on including the $a_2$-meson contribution
 in the dispersive analysis of the radiative decay
 $\eta\to \pi^+ \pi^- \gamma$~\cite{Kubis_2015}, which is connected
 to the isovector contributions of the $\eta \to \gamma \gamma^*$ TFF.
 This resulted in a better control of $F_{\eta\gamma^*\gamma}$ calculations
 and a better consistency of these calculations with those for
 $F_{\eta'\gamma^*\gamma}$.  Also in Ref.~\cite{Xiao_2015},
 a better consistency was reached between the single off-shell form factor
 $F_{\eta\gamma^*\gamma}$ and the double off-shell form factor
 $F_{\eta\gamma^*\gamma^*}$, an accurate model-independent determination of which
 would be an important step towards a reliable evaluation of the HLbL
 scattering contribution to $(g-2)_\mu$.

\subsection{Dalitz decays of $\omega$}
\label{Intro-OmegaDecay}
 The situation is quite different for
 the $\omega \to \pi^0 \gamma^* \to \pi^0 \ell^+\ell^-$ decay.
 The experimental data are available only for
 the $\omega \to \pi^0 \mu^+\mu^-$ decay, showing fair consistency with each other.
 However, the existing theoretical approaches, which successfully reproduce most recent TFF
 data available for $\eta$ and other light mesons in different momentum-transfer
 regions, cannot describe the TFF data based on
 the $\omega \to \pi^0 \gamma^* \to \pi^0 \mu^+\mu^-$ decay at large $m(\mu^+\mu^-)$.
 
 The pioneering measurement of $\omega \to \pi^0 \mu^+\mu^-$,
 $\Lambda^{-2}_{\omega\pi^0}=(2.36\pm0.21_{\rm{tot}})$~GeV$^{-2}$, by 
 Lepton-G~\cite{Lepton_G_omega}, made a few decades ago, was based on
 $60\pm 9$ observed events. The level of background events, which comprised
 11\% from nonresonant sources and 3\% from the $\omega \to \pi^0 \pi^+ \pi^-$ decay
 (with charged pions decaying into muons) and the $\rho\to\pi^0\mu^+\mu^-$ decay,
 could be significant at high $m(\mu^+\mu^-)$, where the theoretical predictions
 were not able to describe the Lepton-G data.
 Most recent measurements by the NA60 experiment in peripheral In--In data~\cite{NA60_2009},
 $\Lambda^{-2}_{\omega\pi^0}=(2.24\pm 0.06_{\mathrm{stat}}\pm 0.02_{\mathrm{syst}})$~GeV$^{-2}$
 from $3\cdot 10^3$ $\omega \to \pi^0 \mu^+\mu^-$ decays,
 and in p--A collisions~\cite{NA60_2016},
 $\Lambda^{-2}_{\omega\pi^0}=(2.223\pm 0.026_{\mathrm{stat}}\pm 0.037_{\mathrm{syst}})$~GeV$^{-2}$,
 were based on measuring the entire spectrum of the $\mu^+\mu^-$ invariant masses,
 without detecting any neutral final-state particles.
 All contributions, except $\eta\to\mu^+\mu^-\gamma$, $\omega\to\pi^0\mu^+\mu^-$,
 and $\rho\to\mu^+\mu^-$, were subtracted from this spectrum.
 The acceptance-corrected spectrum was then fitted with these three contributions.
 According to a more scrupulous analysis of p--A collisions, involving also much higher
 statistics than peripheral In--In data, all possible systematic
 uncertainties were very carefully taken into account.
 Although these latest $|F_{\omega\pi^0}|^2$ results were slightly lower than
 from peripheral In--In data, they confirmed once again the discrepancy with
 the available predictions in the vicinity of the kinematic limit.

 In Refs.~\cite{TL10,Ter12}, the calculations of the $\omega\pi^0$ TFF were based
 on a chiral Lagrangian approach; this
 included light vector mesons and Goldstone bosons to calculate the decays of
 light vector mesons into a pseudoscalar meson and a dilepton in leading order.
 Recent calculations based on dispersion theory were presented
 in Refs.~\cite{Schneider_2012,Danilkin_2015}. In Ref.~\cite{Schneider_2012},
 these calculations and their theoretical uncertainties relied on a previous dispersive
 analysis~\cite{Niecknig_2012} of the corresponding three-pion decays and
 the pion vector form factor. 
 In Ref.~\cite{Danilkin_2015}, a similar dispersive analysis is performed
 for the same three-pion decays ($\omega/\phi\to\pi^+\pi^-\pi^0$)
 with an additional parametrization of the inelastic contributions
 by a power series in a suitably chosen conformal variable that took
 into account the change in the analytical behavior of the amplitude.
 As a further application of this formalism, the e/m TFFs
 of $\omega/\phi \to \pi^0 \gamma^*$ were also computed.

 Motivated by the discrepancies between the theoretical calculations of
 the $\omega\pi^0$ TFF and the experimental data, a further investigation
 of this form factor was made by using analyticity and unitarity in
 a framework known as the method of unitarity bounds~\cite{Caprini_2014}.
 The results for the upper and lower bounds on $|F_{\omega\pi^0}|^2$ in the elastic region
 provided a significant check on those obtained with standard dispersion relations,
 confirming the existence of a disagreement with experimental data in the $q$ region around 0.6~GeV.
 Other tests of the consistency of the $\omega\pi^0$ TFF with unitarity and analyticity
 were recently reported in Ref.~\cite{Caprini_2015}.
 A dispersive analysis of the $\omega\pi^0$ e/m TFF described in this work used
 as input the discontinuity provided by unitarity below the $\omega\pi^0$ threshold
 and, for the first time, included experimental data on the modulus
 measured from $e^+e^-\to \omega\pi^0$ at higher energies. That analysis also
 confirmed the discrepancy between the experimental data and the theoretical
 calculation of the $\omega\pi^0$ TFF in this $q$ region.

\subsection{Dalitz decays with A2}
\label{Intro-A2}
 Compared to the $\omega \to \pi^0 \mu^+\mu^-$ decay, the advantage of measuring
 $\omega\to \pi^0e^+e^-$ would be in giving access
 to the TFF energy dependence at low momentum transfer, which is important
 for data-driven approaches calculating the corresponding rare decays
 and the HLbL contribution to $(g-2)_\mu$.
 The capability of the A2 experimental setup to measure Dalitz decays
 was demonstrated in Refs.~\cite{eta_tff_a2_2014,eta_tff_a2_2011} for 
 $\eta \to e^+e^-\gamma$.
 Measuring $\omega\to \pi^0e^+e^-$ with the A2 setup
 is more challenging because of a much smaller signal, compared to background
 contributions. Nonresonant contributions, like $\pi^0\pi^0$ and $\pi^0\eta$
 final states can cause the same number of electromagnetic showers
 as the $\pi^0e^+e^-$ final state. Also, both $\pi^0$ and $\eta$ have
 their own $e^+e^-\gamma$ decay modes, resulting in $e^+e^-$ pairs that can be
 detected along with $\pi^0$, if the photon from the former decay is not detected.
 The $\omega \to \pi^0 \pi^+ \pi^-$ decay, which has a branching ratio
 one order of magnitude larger than that for $\omega\to \pi^0\gamma$,
 can mimic the $\pi^0e^+e^-$ final state when both charged pions
 deposit their total energy due to nuclear interactions in an
 electromagnetic calorimeter. Because of the smallness of
 the $\eta \to \pi^+\pi^-\gamma$ branching ratio, such a problem does not
 exist for the $\eta \to e^+e^-\gamma$ decay.
 Another decay, $\eta \to \pi^+\pi^-\pi^0$, with a larger branching ratio,
 cannot mimic an $\eta \to e^+e^-\gamma$ peak with one final-state photon being undetected. 
 Thus, the background situation requires a more sophisticated analysis
 for measuring $\omega\to \pi^0e^+e^-$ than is needed for $\eta \to e^+e^-\gamma$.
 To improve the statistical accuracy, two sets of A2 data from 2007 and 2009 were
 analyzed independently, and their results were combined together.
 The same technique was tested with $\eta \to e^+e^-\gamma$ events,
 which have much better statistics and less background, in order to
 determine the effect on the systematic uncertainty caused by this more
 sophisticated analysis. Including 2009 data in the present analysis 
 doubled the $\eta \to e^+e^-\gamma$ statistics, compared to the previous analysis
 of only 2007 data~\cite{eta_tff_a2_2014}, and, along with other improvements,
 resulted in a better accuracy of the A2 results for this Dalitz decay.  
    
 The new results for the $\eta$ and $\omega\pi^0$ e/m TFFs
 presented in this paper are based on measuring $\eta \to e^+e^-\gamma$
 and $\omega\to \pi^0e^+e^-$ decays from a total of
 $5.87 \cdot 10^7$ $\eta$ mesons and $2.27 \cdot 10^7$ $\omega$ mesons
 produced in the $\gamma p\to \eta p$ and $\gamma p\to \omega p$ reactions,
 respectively. Previously, the same data sets were used, for instance,
 in a measurement of the $\eta \to \pi^0\gamma\gamma$
 decay~\cite{etapi0gg_a2_2014}. In addition to the increase in
 the experimental statistics, compared to the previous $\eta \to e^+e^-\gamma$
 measurements~\cite{eta_tff_a2_2014,eta_tff_a2_2011} by the A2 Collaboration,
 the present TFF results include systematic uncertainties in every individual
 data point. This allows a more fair comparison of the data with theoretical
 calculations, especially those calculations which do not follow the VMD pole
 approximation, typically used to fit the data in experimental analyses.
 Data-driven approaches would also prefer data points with total
 uncertainties, rather than measurements with
 the systematic uncertainties given only for the slope-parameter values.   
 As in the case of the previous measurements, radiative corrections to the QED
 differential decay rate of the $\eta$ and $\omega$ Dalitz decays were not taken
 into account in the present work because their precise magnitude had not been
 calculated, but possible systematic uncertainties due to those corrections are
 discussed further in the text.

\section{Experimental setup}
\label{sec:Setup}

The processes $\gamma p\to \eta p \to e^+e^-\gamma p$
and $\gamma p\to \omega p \to \pi^0e^+e^- p$
were measured by using the Crystal Ball (CB)~\cite{CB}
as a central calorimeter and TAPS~\cite{TAPS,TAPS2}
as a forward calorimeter. These detectors were
installed in the energy-tagged bremsstrahlung photon beam of
the Mainz Microtron (MAMI)~\cite{MAMI,MAMIC}. 
The photon energies were determined by using
the Glasgow--Mainz tagging spectrometer~\cite{TAGGER,TAGGER1,TAGGER2}.

The CB detector is a sphere consisting of 672
optically isolated NaI(Tl) crystals, shaped as
truncated triangular pyramids, which point toward
the center of the sphere. The crystals are arranged in two
hemispheres that cover 93\% of $4\pi$, sitting
outside a central spherical cavity with a radius of
25~cm, which holds the target and inner
detectors. In this experiment, TAPS was
arranged in a plane consisting of 384 BaF$_2$
counters of hexagonal cross section.
It was installed 1.5~m downstream of the CB center
and covered the full azimuthal range for polar angles
from $1^\circ$ to $20^\circ$.
More details on the energy and angular resolution of the CB and TAPS
are given in Refs.~\cite{slopemamic,etamamic}.

 The present measurement used electron beams
 with energies of 1508 and 1557 MeV from the Mainz Microtron, MAMI-C~\cite{MAMIC}.
 The data with the 1508-MeV beam were taken in 2007 (Run-I)
 and those with the 1557-MeV beam in 2009 (Run-II).
 Bremsstrahlung photons, produced by the beam electrons
 in a 10-$\mu$m Cu radiator and collimated by a 4-mm-diameter Pb collimator,
 were incident on a liquid hydrogen (LH$_2$) target located
 in the center of the CB. The LH$_2$ target was 5-cm and 10-cm long
 in Run-I and Run-II, respectively.
 The total amount of material around the LH$_2$ target,
 including the Kapton cell and the 1-mm-thick carbon-fiber beamline,
 was equivalent to 0.8\% of a radiation length $X_0$.
 In the present measurement, it was essential to keep the material budget
 as low as possible to minimize the background from $\eta \to \gamma\gamma$
 and $\omega\to \pi^0\gamma$ decays with conversion of the photons into $e^+e^-$ pairs.
\begin{figure}
\includegraphics[width=7.cm,height=8.cm,bbllx=0.cm,bblly=0.cm,bburx=10.5cm,bbury=12.cm]{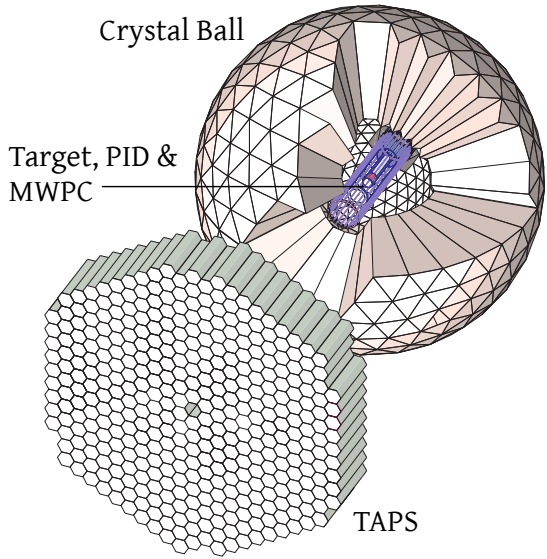}
\caption{(Color online)
  A general sketch of the Crystal Ball, TAPS, and particle identification (PID) detectors.
}
 \label{fig:cb_taps_pid} 
\end{figure}

 The target was surrounded by a Particle IDentification
 (PID) detector~\cite{PID} used to distinguish between charged and
 neutral particles. It is made of 24 scintillator bars
 (50 cm long, 4 mm thick) arranged as a cylinder with a radius of 12 cm.
 A general sketch of the CB, TAPS, and PID is shown
 in Fig.~\ref{fig:cb_taps_pid}.
 A multiwire proportional chamber, MWPC, also shown in this figure
 (which consists of two cylindrical MWPCs inside each other),
 was not installed during Run-I and was not used during Run-II
 as it could not operate in the high photon flux used in this experiment.

 In Run-I, the energies of the incident photons were analyzed
 up to 1402~MeV by detecting the postbremsstrahlung electrons
 in the Glasgow tagged-photon spectrometer
 (Glasgow tagger)~\cite{TAGGER,TAGGER1,TAGGER2},
 and up to 1448~MeV in Run-II.
 The uncertainty in the energy of the tagged photons is mainly determined
 by the segmentation of the tagger focal-plane detector in combination with
 the energy of the MAMI electron beam used in the experiments.
 Increasing the MAMI energy increases the energy range covered
 by the spectrometer and also has the corresponding effect on the uncertainty
 in $E_\gamma$. For both the MAMI energy settings of 1508 and 1557~MeV,
 this uncertainty was about $\pm 2$~MeV.
 More details on the tagger energy calibration and uncertainties
 in the energies can be found in Ref.~\cite{EtaMassA2}.

 The experimental trigger in Run-I required the total energy deposited in the CB
 to exceed $\sim$320~MeV and the number of so-called hardware clusters
 in the CB (multiplicity trigger) to be two or more.
 In the trigger, a hardware cluster in the CB was a block of 16
 adjacent crystals in which at least one crystal had an energy
 deposit larger than 30 MeV.
 Depending on the data-taking period, events with a cluster multiplicity
 of two were prescaled with different rates.
 TAPS was not included in the multiplicity trigger for these experiments.
 In Run-II, the trigger on the total energy
 in the CB was increased to $\sim$340~MeV, and the multiplicity
 trigger required $\ge3$ hardware clusters in the CB.

\section{Data handling}
\label{sec:Data}
\subsection{Selection of candidate events}
\label{subsec:Data-I}
 To search for a signal from $\eta \to e^+e^-\gamma$ decays, 
 candidates for the process $\gamma p\to e^+e^-\gamma p$
 were extracted from events having three or four clusters reconstructed
 by a software analysis in the CB and TAPS together.
 The offline cluster algorithm was optimized for finding
 a group of adjacent crystals in which the energy was deposited
 by a single-photon e/m shower. This algorithm
 works well for $e^{+/-}$, which also produce
 e/m showers in the CB and TAPS, and for proton clusters.
 The software threshold for the cluster energy was chosen to be 12 MeV.
 For the $\gamma p\to e^+e^-\gamma p$ candidates, 
 the three-cluster events were analyzed assuming that the final-state
 proton was not detected. The fraction of such $\eta \to e^+e^-\gamma$
 decays was only about 20\% from the total. 
 Compared to the previous analysis of $\eta \to e^+e^-\gamma$,
 reported in Ref.~\cite{eta_tff_a2_2014}, there were some improvements
 that resulted in a more reliable particle identification and in fewer 
 sources of systematic uncertainties. Such improvements are discussed
 later in the text, including, for instance, the PID $dE/dx$ analysis
 for particle identification and adding the angular dependence of
 the virtual-photon decay in the Monte Carlo (MC) event generator for a more reliable
 acceptance determination. 

 To search for a signal from $\omega \to \pi^0e^+e^-$ decays, 
 candidates for the process $\gamma p\to \pi^0e^+e^- p\to \gamma\gamma e^+e^- p$
 were extracted from the analysis of events having
 five clusters (four from the photons and one from the proton)
 reconstructed in the CB and TAPS together.
 Four-cluster events, with only four photons detected,
 were neglected in the analysis because
 the proton information missing for such events in the analysis
 resulted in a much stronger background.
 In addition, as shown in Ref.~\cite{a2_omegap_2015},
 the fraction of  $\gamma p\to \omega p\to \pi^0\gamma p$ events without
 the detected proton was quite small, varying from 2.7\% at the reaction
 threshold to 7.6\% at the highest energy of the present experiments.

 The selection of candidate events and the reconstruction of the reaction
 kinematics was based on the kinematic-fit technique.
 Details of the kinematic-fit parametrization of the detector
 information and resolutions are given in Ref.~\cite{slopemamic}.
 Because the three-cluster sample, in which there are good
 $\gamma p\to \eta p \to e^+e^-\gamma p$ events without the outgoing
 proton detected, was mostly dominated by $\gamma p\to \pi^0p\to \gamma\gamma p$
 and $\gamma p\to \eta p\to \gamma\gamma p$ events, the corresponding
 kinematic-fit hypotheses were tested first. Then all events for which
 the confidence level (CL) to be $\gamma p\to \pi^0p$ or $\gamma p\to \eta p$
 was greater than $10^{-5}$ were discarded from further analysis. It was
 checked that such a preselection practically does not cause any losses of
 $\eta \to e^+e^-\gamma$ decays (which are $<1\%$), but rejects a significant
 background from two-photon final states.
 Because e/m showers from electrons and positrons are
 very similar to those of photons, 
 the hypothesis $\gamma p \to 3\gamma p$ was tested to identify
 the $\gamma p\to e^+e^-\gamma p$ candidates.
 To identify $\omega \to \pi^0e^+e^-$ candidates, two hypotheses,
 $\gamma p\to 4\gamma p$ and
 $\gamma p\to \pi^0\gamma\gamma p\to 4\gamma p$, were tested.
 The events that satisfied these hypotheses with the CL greater
 than 1\% were accepted for further analysis. The kinematic-fit output
 was used to reconstruct the kinematics of the outgoing particles.
 In this output, there was no separation between e/m showers
 caused by the outgoing photon, electron, or positron.  
 Because the main purpose of the experiments was to measure
 the $\eta \to e^+e^-\gamma$ and $\omega \to \pi^0e^+e^-$ decay rates
 as a function of the invariant mass $m(e^+e^-)$, the next step
 in the analysis was the separation of $e^+e^-$ pairs from
 final-state photons. This procedure was optimized by using a MC
 simulation of the processes $\gamma p\to \eta p \to e^+e^-\gamma p$
 and $\gamma p\to \omega p\to \pi^0e^+e^- p\to \gamma\gamma e^+e^- p$.

\subsection{Monte Carlo simulations}
\label{subsec:Data-II}
 Those MC simulations were made to be as similar as possible to the real events
 to minimize the systematic uncertainties in the determination of
 experimental acceptances and to properly measure the energy dependence of the TFFs.
 To reproduce the experimental yield of $\eta$ and $\omega$ mesons and
 their angular distributions as a function of the incident-photon
 energy, both the $\gamma p\to \eta p$ and $\gamma p\to \omega p$ reactions
 were generated according to the numbers of the corresponding events and their
 angular distributions measured in the same experiment~\cite{etamamic,a2_omegap_2015}.
 The $\eta \to e^+e^-\gamma$ decays were generated according to Eq.~(\ref{eqn:dgdm_eta}),
 with the phase-space term removed and with $\Lambda^{-2}_{\eta} = 1.95$~GeV$^{-2}$
 from previous experiments~\cite{NA60_2009,eta_tff_a2_2014}.
 The generation of the $\omega \to \pi^0e^+e^-$ decays were made in two steps.
 To reproduce the energy dependence of the $\omega$
 decay width near the production threshold, the reaction
 $\gamma p\to \pi^0e^+e^-p$ was generated first according to phase space.
 Then, the invariant mass $m(\pi^0e^+e^-)$ was folded with the Breit-Wigner
 (BW) function, with the parameters taken for the $\omega$ meson
 from the Review of Particle Physics (RPP)~\cite{PDG}.
 This approach allowed one to properly reproduce
 the folding of the BW shape with phase space.
 Next, the invariant mass $m(e^+e^-)$ was folded to follow
 Eq.~(\ref{eqn:dgdm_omega}) with $\Lambda^{-2}_{\omega\pi^0} = 2.24$~GeV$^{-2}$~\cite{NA60_2009}.  
 The angular dependence of the virtual photon decaying into a lepton pair
 was generated according to Eq.~(\ref{eqn:dtheta}),
 for both $\eta\to \gamma^*\gamma \to \ell^+\ell^-\gamma$ and
 $\omega\to \pi^0\gamma^* \to \pi^0\ell^+\ell^-$.

 Possible background processes were also studied by using MC simulations.
 The reaction $\gamma p\to \eta p$ was simulated for several other decay modes
 of the $\eta$ meson to check if they could mimic a peak
 from the $\eta \to e^+e^-\gamma$ signal.
 Such MC simulations were made for the $\eta\to\gamma\gamma$, $\eta\to\pi^0\pi^0\pi^0$,
 $\eta\to\pi^+\pi^-\pi^0$, and  $\eta\to\pi^+\pi^-\gamma$ decays.
 The yield and the production angular distributions of
 all $\gamma p\to \eta p$ simulations were generated in the same way as 
 for the process $\gamma p\to \eta p \to e^+e^-\gamma p$. 
 In contrast to the $\eta \to e^+e^-\gamma$ decay, all other $\eta$ decays
 were generated according to phase space.
 The major background under the peak from $\eta \to e^+e^-\gamma$ decays
 was found to be from the reaction $\gamma p\to \pi^0\pi^0p$.
 The MC simulation of this reaction was carried out in the same way as 
 reported in Ref.~\cite{p2pi0mamic}.
 Although this background is smooth in the region of the $\eta$ mass and 
 cannot mimic an $\eta \to e^+e^-\gamma$ peak, its MC simulation was
 used to optimize the signal-to-background ratio and to
 parametrize the background under the signal.

 A similar study was also made for the $\omega \to \pi^0e^+e^-$ decay.
 The reaction $\gamma p\to \omega p$ was simulated for $\omega \to \pi^0\gamma$,
 with both the $\gamma\gamma$ and $\gamma e^+e^-$ decay modes of the $\pi^0$,
 and for $\omega\to \pi^+\pi^-\pi^0$ decays.
 The $\omega$ decay width was reproduced by folding $m(\pi^0\gamma)$
 and $m(\pi^+\pi^-\pi^0)$ in the processes $\gamma p\to \pi^0\gamma p$ and
 $\gamma p\to \pi^+\pi^-\pi^0 p$ with the BW function having
 $\omega$ parameters from the RPP~\cite{PDG}. The Dalitz decay of
 the $\pi^0$ in $\omega \to \pi^0\gamma$ was generated according to its
 pure QED dependence. Additionally to the $\gamma p\to \pi^0\pi^0p$
 background, the simulation of which was also needed for $\eta \to e^+e^-\gamma$,
 a study of the $\gamma p\to \pi^0\eta p$ background was made via
 its simulation. 

 For all reactions, the simulated events
 were propagated through a {\sc GEANT} (version 3.21) simulation of the experimental
 setup. To reproduce the resolutions observed in the experimental data, the {\sc GEANT}
 output (energy and timing) was subject to additional smearing, thus
 allowing both the simulated and experimental data to be analyzed in the same way.
 Matching the energy resolution between the experimental and MC events
 was achieved by adjusting the invariant-mass resolutions,
 the kinematic-fit stretch functions (or pulls), and probability
 distributions. Such an adjustment was based on the analysis of the
 same data sets for reactions having almost no background
 from other physical reactions
 (namely, $\gamma p\to \pi^0 p$, $\gamma p\to \eta p\to \gamma\gamma p$,
 and $\gamma p\to \eta p\to 3\pi^0p$~\cite{slopemamic}).
 The simulated events were also tested to check whether they passed
 the trigger requirements.
 
\subsection{Identifying $e^+e^-$ pairs and suppressing backgrounds}
\label{subsec:Data-III}
 The PID detector was used to identify the final-state $e^+e^-$ pair
 (the detection efficiency for $e^{+/-}$ in the PID is close to 100\%)
 in the events initially selected as $\gamma p \to 3\gamma p$ and
 $\gamma p \to 4\gamma p$ candidates.
 The $\gamma p\to \pi^0\gamma\gamma p\to 4\gamma p$ hypothesis was needed 
 for selecting only $\gamma p\to \pi^0e^+e^-p$ candidates from the five-cluster
 events. Because, with respect to the LH$_2$ target, the PID provides a full
 coverage only for the CB crystals, events with at least three e/m showers in the CB
 were selected for further analysis, allowing one e/m shower to be detected
 in TAPS for $\gamma p\to \pi^0e^+e^-p$ candidates, and requiring the electron
 and positron to be detected in the CB.
 Requiring at least three e/m showers in the CB also made almost all selected events
 pass the trigger requirements on both the total energy and the multiplicity.
 The identification of $e^{+/-}$ in the CB was based on a correlation
 between the $\phi$ angles of fired PID elements with the angles
 of e/m showers in the calorimeter.
 The MC simulations of $\gamma p\to \eta p \to e^+e^-\gamma p$ 
 and $\gamma p\to \omega p\to \pi^0e^+e^- p\to \gamma\gamma e^+e^- p$ were used
 to optimize this procedure, minimizing a probability of misidentification
 of $e^{+/-}$ with the final-state photons. This procedure is optimized with respect
 to how close the $\phi$ angle of an e/m shower in the CB should be to
 the corresponding angle of a fired PID element to be considered
 as $e^{+/-}$, and how far it should be to be considered as a photon.
 This decreases the efficiency in selecting true events for which the $\phi$ angle of
 the electron or the positron is close to the photon $\phi$ angle.

 The analysis of the MC simulations of possible background reactions
 for $\eta \to e^+e^-\gamma$ revealed that only the process
 $\gamma p\to \eta p \to \gamma\gamma p$ could mimic $\eta \to e^+e^-\gamma$ events.
 This occurs mostly when one of the final-state photons converts
 into an $e^+e^-$ pair in the material between the production vertex and
 the NaI(Tl) surface. Because the opening angle between such electrons and positrons
 is typically very small, this background can be significantly suppressed
 by requiring that $e^+$ and $e^-$ were identified by different PID elements.
 However, such a requirement also decreases the detection efficiency for
 actual $\eta \to e^+e^-\gamma$ events, especially at low invariant masses $m(e^+e^-)$.
 In further analysis of $\eta \to e^+e^-\gamma$ events, both the options, with larger
 and smaller background remaining from $\eta\to\gamma\gamma$, were tested.
\begin{figure*}
\includegraphics[width=0.93\textwidth]{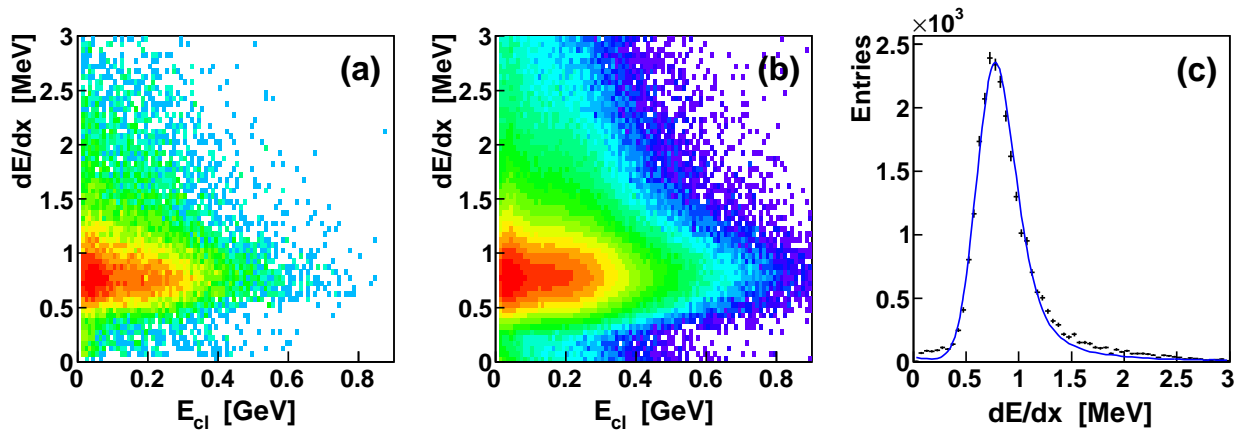}
\caption{ (Color online)
 Comparison of the $e^{+/-}$ $dE/dx$ of the PID for experimental
 $\eta \to e^+e^-\gamma$ decays and their MC simulation.
 The two-dimensional density distribution (with logarithmic scale
 along plot axis $z$) for the $e^{+/-}$ $dE/dx$
 of the PID versus the energy of the corresponding clusters in the CB
 is shown in (a) for the experimental data and in (b) for the MC simulation.
 The $e^{+/-}$ $dE/dx$ distributions for the experimental data (crosses)
 and the MC simulation (blue solid line) are compared in (c).
}
 \label{fig:etaeeg_pid_dedx} 
\end{figure*}
\begin{figure*}
\includegraphics[width=1.\textwidth]{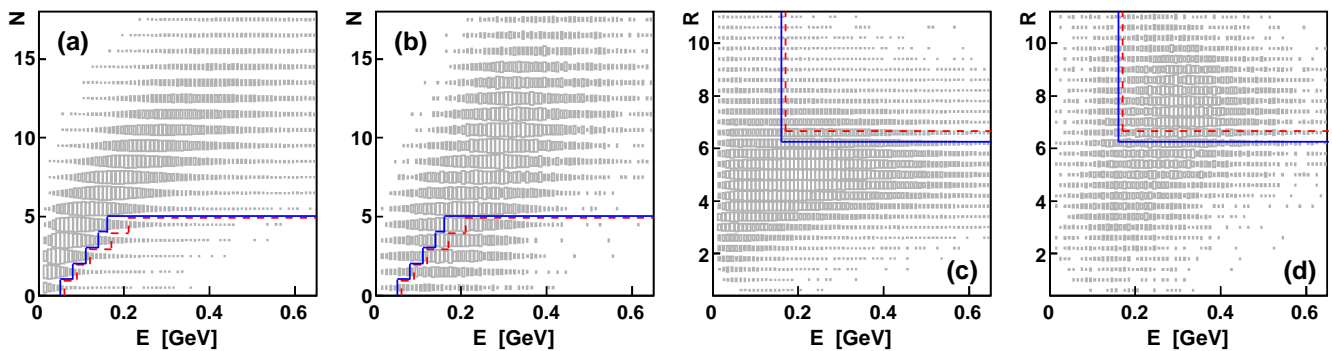}
\caption{ (Color online)
 (a), (c) Two-dimensional density distributions for events selected
 as the $\gamma p\to \pi^0e^+e^- p$ candidates obtained from
 the MC simulations of $\omega\to\pi^0e^+e^-$ decays
 and (b), (d) $\omega\to \pi^0\pi^+\pi^-$ decays causing
 background in the vicinity of the $\omega$ mass.
 Distributions (a) and (b) plot the number $N$ of crystals forming the
 two clusters ascribed to $e^{+/-}$ versus the cluster energies.
 Distributions (c) and (d) plot the effective radius $R$ of the
 two clusters ascribed to $e^{+/-}$ versus the cluster energies.
 The cuts tested for suppressing the $\omega\to \pi^0\pi^+\pi^-$ background
 are shown by red dashed lines for looser cuts and by blue solid lines
 for tighter cuts. The cuts on $N$ discard all events for which at least one
 of the $e^{+/-}$ clusters has an energy larger than the values shown by
 the cut lines. The cuts on effective radius discard all events for which
 at least one of the $e^{+/-}$ clusters has $R$ larger than the values
 shown by the cut lines.
}
 \label{fig:omega_pi0ee_pi_cuts} 
\end{figure*}

 Similarly, the process $\gamma p\to \omega p \to \pi^0\gamma p \to 3\gamma p$
 can mimic $\omega\to\pi^0e^+e^-$ events via converting a final-state photon
 into an $e^+e^-$ pair. Another source of actual $\omega\to \gamma\gamma e^+e^-$
 events comes from $\omega\to\pi^0\gamma$ decays with the Dalitz decay of $\pi^0$.
 Because of the QED dependence of this decay, it dominates at low masses $m(e^+e^-)$
 and can be suppressed by requiring that $e^+$ and $e^-$ were identified
 by different PID elements. Further reduction of this background can be
 achieved by requiring the two final-state photons to be from the $\pi^0$ decay.

 Other background sources that should be significantly suppressed in the analysis
 of $\omega\to\pi^0e^+e^-$ events are the processes $\gamma p\to \pi^0\pi^0p$ and
 $\gamma p\to \pi^0\eta p$, with the $\eta$ meson decaying into two photons
 or into $e^+e^-\gamma$. In the case of the two-photon decay, both the photons can convert
 before or inside the PID, mimicking an $e^+e^-$ pair. In the case of the $e^+e^-\gamma$
 decay, $e^+e^-$ pairs with very low invariant masses often hit the same PID element and
 are reconstructed as one cluster in the CB.
 If the photon from the same decay converts before or inside the PID, such an event
 could be identified as a $\pi^0e^+e^-$ final state. Similarly, the process
 $\gamma p\to \pi^0\pi^0p$ can mimic $\gamma p\to \pi^0e^+e^-p$ events.
 Without suppressing background from $\gamma p\to \pi^0\pi^0p$ and
 $\gamma p\to \pi^0\eta p$, the signal
 from $\omega\to\pi^0e^+e^-$ would be comparable with the statistical fluctuations
 of the background events, preventing the measurement of the TFF at $m(e^+e^-)$ close
 to the $\pi^0$ and $\eta$ masses, with the $\eta$-mass region being especially
 important for the $\omega\pi^0$ TFF.

 The suppression of background from $\gamma p\to \pi^0\pi^0p$ and $\gamma p\to \pi^0\eta p$
 was based on the analysis of energy losses, $dE/dx$, in the PID elements.
 According to the MC simulations of these backgrounds, many photons produce energy losses
 that are significantly smaller than $dE/dx$ from a single $e^{+/-}$, and the $e^+e^-$ pairs
 reconstructed as one cluster in the CB result in a double-magnitude PID signal,
 compared to a single $e^{+/-}$. To reflect the actual differential energy deposit
 $dE/dx$ in the PID, the energy signal from each element,
 ascribed to either $e^+$ or $e^-$, was multiplied by
 the sine of the polar angle of the corresponding particle,
 the magnitude of which is taken from the kinematic-fit output.
 All PID elements were calibrated so that the $e^{+/-}$ peak position matched
 the corresponding peak in the MC simulation.
 To reproduce the actual energy resolution of the PID with the MC simulation,  
 the {\sc GEANT} output for PID energies was subject to additional smearing,
 allowing the $e^{+/-}$ selection with $dE/dx$ cuts to be very similar for
 the experimental data and MC. 
 The PID energy resolution in the MC simulations was adjusted to match
 the experimental $dE/dx$ spectra for the $e^{+/-}$ particles
 produced in $\eta \to e^+e^-\gamma$ decays with $m(e^+e^-)$ below
 the $\pi^0$ mass, the range in which these decays can be selected with
 very small background, especially if the final-state proton is detected
 (this will be illustrated further in the text).
 The same sample was used to check possible
 systematic uncertainties due to losses of good events while applying $dE/dx$ cuts
 to suppress background from $\gamma p\to \pi^0\pi^0p$ and $\gamma p\to \pi^0\eta p$.

 The experimental $dE/dx$ resolution of the PID for $e^{+/-}$ and the comparison of
 it with the MC simulation is illustrated in Fig.~\ref{fig:etaeeg_pid_dedx}.
 Figures~\ref{fig:etaeeg_pid_dedx}(a) and (b) compare the experimental and MC-simulation
 plots of the $e^{+/-}$ $dE/dx$ of the PID versus the energy of the corresponding
 clusters in the CB. As seen, there is no $dE/dx$ dependence of $e^{+/-}$
 on their energy in the CB, and applying cuts just on a $dE/dx$ value is sufficient
 for suppressing background from $\gamma p\to \pi^0\pi^0p$ and $\gamma p\to \pi^0\eta p$.
 The comparison of the experimental $e^{+/-}$ $dE/dx$ distributions with
 the MC simulation is depicted in Fig.~\ref{fig:etaeeg_pid_dedx}(c).
 A small difference in the tails of the $e^{+/-}$ peak can be explained
 by some background remaining in the experimental spectrum.
 Typical PID cuts, which were tested, varied from requiring $dE/dx<2.7$~MeV to
 $dE/dx<1.2$~MeV to suppress background events, showing no systematic effects
 in the final results. 

 The $\omega \to \pi^0 \pi^+ \pi^-$ decay
 can mimic the $\pi^0e^+e^-$ final state when both charged pions
 deposit their total energy due to nuclear interactions in the CB.
 The probability of such events is quite
 low, but the branching ratio for $\omega \to \pi^0 \pi^+ \pi^-$ is a factor
 $\sim2\times10^3$ greater than for $\omega\to\pi^0e^+e^-$. The suppression
 of the $\omega \to \pi^0 \pi^+ \pi^-$ background to a level
 negligible for $\omega\to\pi^0e^+e^-$ events typically requires
 a combination of a few selection criteria.   
 The energy resolution of the PID is not sufficient
 to efficiently separate $\pi^{+/-}$ from $e^{+/-}$ by the $dE/dx$
 method. Most of the background events from $\omega \to \pi^0 \pi^+ \pi^-$
 decays have a low probability for $\gamma p\to \pi^0\gamma\gamma p\to 4\gamma p$,
 and the position of the event vertex along the beam direction
 ($z$ axis), reconstructed by the kinematic
 fit, is strongly shifted in the downstream direction. Such a shift in $z$ is
 caused by an attempt by the kinematic fit to compensate for an imbalance in
 energy conservation by changing significantly the polar angles of the outgoing
 particles, which is only possible by moving the event vertex along the beam
 direction. Accordingly, applying cuts on the
 kinematic-fit CL and the vertex coordinate $z$ mostly rejects
 $\omega \to \pi^0 \pi^+ \pi^-$ events with cluster energies of $\pi^{+/-}$
 below their total energies. Typically, such events are reconstructed by
 the kinematic fit with invariant masses $m(\pi^0e^+e^-)$ below the mass
 of the $\omega$ meson. Further suppression of the $\omega \to \pi^0\pi^+\pi^-$
 background events remaining in the vicinity of the $\omega$ mass can be achieved
 by using differences in features of e/m and nuclear-interaction showers in the CB.
 As observed from MC simulations of $\omega\to\pi^0e^+e^-$ and $\omega\to \pi^0\pi^+\pi^-$
 decays, nuclear-interaction showers at lower energies typically have a smaller
 multiplicity of the crystals forming a cluster. At higher energies,
 nuclear-interaction showers from $\pi^{+/-}$ typically spread more
 widely than e/m showers. Such a spread can be evaluated via the cluster
 effective radius. For the CB, the effective radius $R$ of a cluster containing
 $k$ crystals with energy $E_i$ deposited in crystal $i$ can be defined as
\begin{equation}
 R = (\sum^k_i{E_i (\Delta r_i)^2}/\sum^k_i{E_i})^{1/2}~,
\label{eqn:rad}
\end{equation}
 where $\Delta r_i$ is the opening angle (in degrees) between the cluster
 direction (as determined by the cluster algorithm) and the crystal-axis direction.
 The multiplicity $N$ of the crystals forming a cluster ascribed to $e^{+/-}$
 is shown as a function of the cluster energy in
 Figs.~\ref{fig:omega_pi0ee_pi_cuts}(a) and (b), respectively for
 the MC simulations of $\omega\to\pi^0e^+e^-$ decays and 
 $\omega\to \pi^0\pi^+\pi^-$ decays causing background in the vicinity
 of the $\omega$ mass. Similar distributions for the effective radius $R$
 of clusters ascribed to $e^{+/-}$ are shown in
 Figs.~\ref{fig:omega_pi0ee_pi_cuts}(c) and (d).
 The cuts tested for suppressing the $\omega\to \pi^0\pi^+\pi^-$ background
 are depicted by red dashed lines for looser cuts and by blue solid lines
 for tighter cuts. The cuts on $N$ discard all events for which at least one
 of the clusters ascribed to $e^{+/-}$ has an energy larger than the value
 shown by the corresponding cut lines. The cuts on effective radius discard
 all events for which at least one of $e^{+/-}$ clusters has $R$ larger
 than the values shown by the cut lines.
 These cuts were optimized to significantly suppress
 the $\omega\to \pi^0\pi^+\pi^-$ background with minimal losses of
 $\omega\to\pi^0e^+e^-$ decays. To make sure that these cuts do not
 cause systematic uncertainties in the $\omega\to\pi^0e^+e^-$ results,
 the same cuts were tested in the analysis of the $\eta \to e^+e^-\gamma$
 decay, which has much better statistics and less background.
\begin{figure}
\includegraphics[width=0.45\textwidth,height=4.35cm,bbllx=1.cm,bblly=.5cm,bburx=19.5cm,bbury=9.cm]{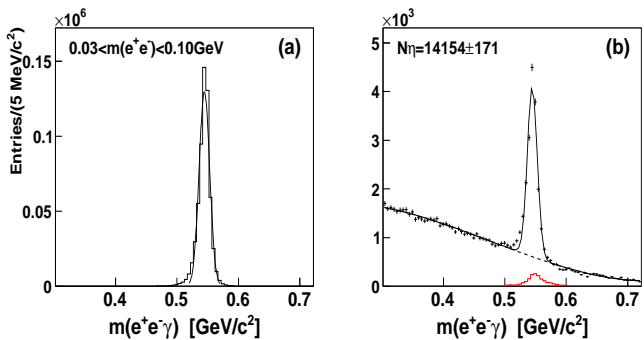}
\caption{ (Color online)
 $m(e^+e^-\gamma)$ invariant-mass distributions obtained for the $m(e^+e^-)$
 range from 30 to 100~MeV/$c^2$ with $\gamma p\to e^+e^-\gamma p$ candidates
 initially selected with the kinematic fit and allowing both $e^+$ and $e^-$
 to be identified with the same PID element:
 (a)~MC simulation of $\gamma p\to \eta p \to e^+e^-\gamma p$ fitted with
     a Gaussian;
 (b)~experimental events (Run-I) after subtracting the random background
     and the remaining background from $\gamma p\to \eta p \to \gamma\gamma p$.
     The distribution for the $\eta \to \gamma\gamma$ background, shown
     by a red solid line, is normalized to the number of subtracted events.
     The experimental distribution is fitted with the sum of a Gaussian
     for the $\eta \to e^+e^-\gamma$ peak and a polynomial of order four
     for the background. The total fit is depicted by a solid line,
     and the dashed line shows the background under the peak. 
}
 \label{fig:eegz34_eta_cth_2007_fit10_m30_100} 
\end{figure}
\begin{figure}
\includegraphics[width=0.45\textwidth,height=4.35cm,bbllx=1.cm,bblly=.5cm,bburx=19.5cm,bbury=9.cm]{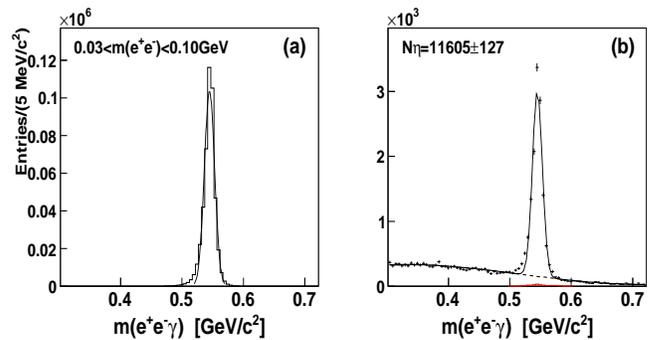}
\caption{ (Color online)
 Same as Fig.~\protect\ref{fig:eegz34_eta_cth_2007_fit10_m30_100},
 but requiring both $e^+$ and $e^-$ to be identified
 by different PID elements.
}
 \label{fig:eegz34_eta_cth_2007_fit0_m30_100} 
\end{figure}
\begin{figure}
\includegraphics[width=0.45\textwidth,height=4.35cm,bbllx=1.cm,bblly=.5cm,bburx=19.5cm,bbury=9.cm]{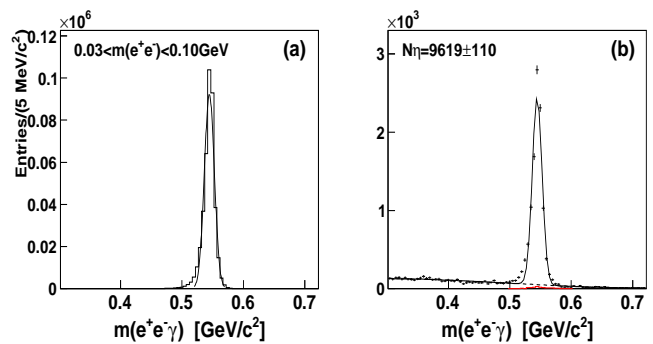}
\caption{ (Color online)
 Same as Fig.~\protect\ref{fig:eegz34_eta_cth_2007_fit0_m30_100}, but after
 applying the $dE/dx$ PID cut selecting only events with the elements having
 an energy deposit corresponding to a single $e^{+/-}$, and requiring
 the final-state proton to be detected.
}
 \label{fig:eegz34_eta_cth_2007_fit1_4_m30_100} 
\end{figure}

 In addition to the background contributions from other physical
 reactions, there are two more background sources.
 The first source comes from interactions of
 incident photons in the windows of the target cell.
 The subtraction of this background from
 experimental spectra is typically based on the 
 analysis of data samples that were taken
 with an empty target. In the present analysis, the empty-target
 background was small and did not feature any visible
 $\eta$ peak in its $m(e^+e^-\gamma)$ spectra
 for the $\gamma p\to e^+e^-\gamma p$ candidates nor
 any $\omega$ peak in its $m(\pi^0e^+e^-)$ spectra
 for the $\gamma p\to \pi^0e^+e^- p$ candidates.
 Another background was caused by random coincidences
 of the tagger counts with the experimental trigger;
 its subtraction was carried out by using 
 event samples for which all coincidences were random
 (see Refs.~\cite{slopemamic,etamamic} for more details).
\begin{figure*}
\includegraphics[width=0.85\textwidth]{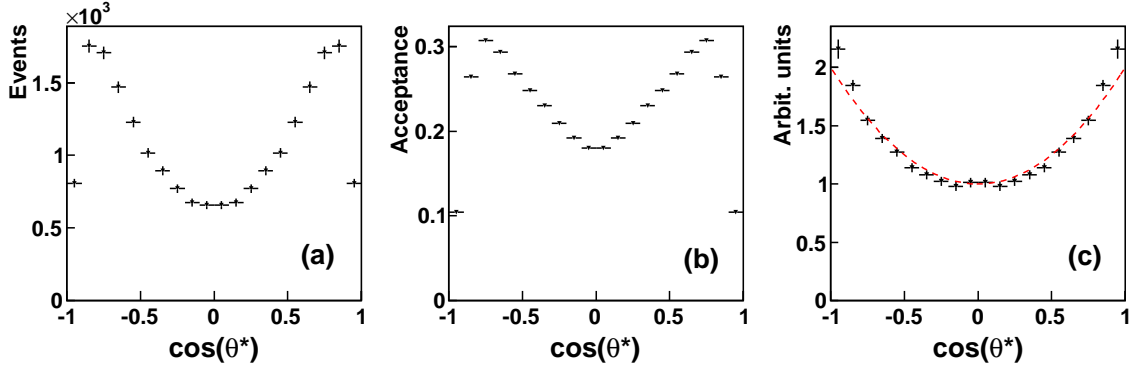}
\caption{ (Color online)
 The $\eta\to \gamma\gamma^* \to \gamma e^+e^-$ angular dependence
 (in the $\eta$ rest frame)
 of the virtual photon decaying into a lepton pair, with $\theta^*$
 being the angle between the direction of one of the leptons in
 the virtual-photon (or the dilepton) rest frame and the direction
 of the dilepton system (which is opposite to the $\gamma$ direction):
 (a) experimental events from the $\eta\to \gamma e^+e^-$ peak;
 (b) angular acceptance based on the MC simulation;
 (c) the experimental spectrum corrected for the acceptance
     and normalized for comparing to the $1 + \cos^2\theta^*$ dependence
     (shown by a red dashed line).
  Because $e^+$ and $e^-$ cannot be separated in the present experiment,
  the angles of both leptons were used, resulting in a symmetric shape
  with respect to $\cos\theta^*=0.$
}
 \label{fig:eta_eeg_cth2e_a2_2007_3x1} 
\end{figure*}

\subsection{Measuring $\eta \to e^+e^-\gamma$ and $\omega\to\pi^0e^+e^-$
 and checking systematic uncertainties}
\label{subsec:Data-IV}
 To measure the $\eta \to e^+e^-\gamma$ and $\omega\to\pi^0e^+e^-$ yield
 as a function of the invariant mass $m(e^+e^-)$, the corresponding
 candidate events were divided into several $m(e^+e^-)$ bins.
 The width of the $m(e^+e^-)$ bins was chosen to be narrower at
 low masses, where the QED dependence results in much higher
 statistics of Dalitz decays, and to be wider at large $m(e^+e^-)$ masses
 with fewer Dalitz decays. Events with $m(e^+e^-)<30$~MeV/$c^2$
 were not analyzed at all, because e/m showers from those $e^+$ and $e^-$
 start to overlap too much in the CB.
 The number of $\eta \to e^+e^-\gamma$ and $\omega\to\pi^0e^+e^-$ decays
 in every $m(e^+e^-)$ bin was determined by fitting
 the experimental $m(e^+e^-\gamma)$ and $m(\pi^0e^+e^-)$ spectra
 with the $\eta$ and $\omega$ peaks rising above a smooth background.
 Possible systematic uncertainties in the results owing to various cuts on
 the kinematic-fit CL, the vertex coordinate $z$, $dE/dx$ of PID,
 the multiplicity $N$ of the crystals forming $e^{+/-}$ clusters, and their
 effective radius $R$ were studied by using enlarged $m(e^+e^-)$ bins,
 allowing greater statistics for such a study.
 The events with $e^+$ and $e^-$ detected with the same PID element
 were analyzed only for the $\eta \to e^+e^-\gamma$ decay.
 For the $\omega\to\pi^0e^+e^-$ decay, such an option resulted
 in too much background in the region of the $\omega$ peak.
\begin{figure}
\includegraphics[width=0.45\textwidth,height=4.35cm,bbllx=1.cm,bblly=.5cm,bburx=19.5cm,bbury=9.cm]{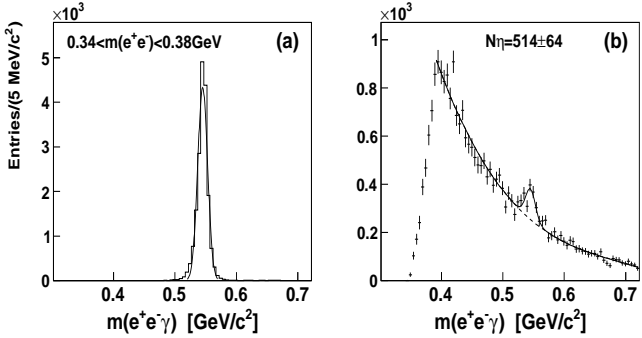}
\caption{
 Same as Fig.~\protect\ref{fig:eegz34_eta_cth_2007_fit10_m30_100}, but for
 $m(e^+e^-)=(360\pm 20)$~MeV/$c^2$. The $\eta \to \gamma\gamma$ background
 is not shown in (b) because it is negligibly small.
}
 \label{fig:eegz34_eta_cth_2007_fit0_m340_380} 
\end{figure}
\begin{figure}
\includegraphics[width=0.45\textwidth,height=4.35cm,bbllx=1.cm,bblly=.5cm,bburx=19.5cm,bbury=9.cm]{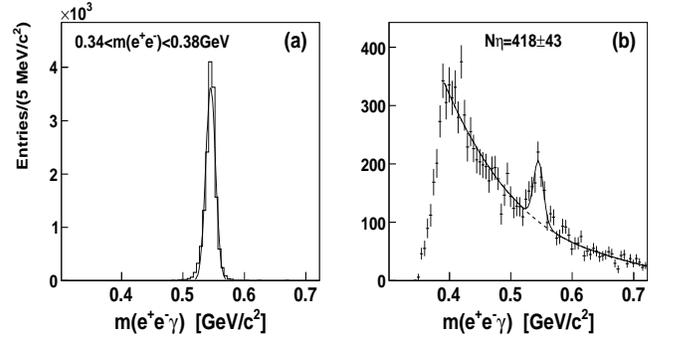}
\caption{
 Same as Fig.~\protect\ref{fig:eegz34_eta_cth_2007_fit0_m340_380}, but after
 applying the softer cuts on $N$ and $R$
 (depicted by red dashed lines in Fig.~\protect\ref{fig:omega_pi0ee_pi_cuts}).
}
 \label{fig:eegz34_eta_cth_2007_fit1_m340_380} 
\end{figure}

\begin{figure}
\includegraphics[width=0.45\textwidth,height=4.35cm,bbllx=1.cm,bblly=.5cm,bburx=19.5cm,bbury=9.cm]{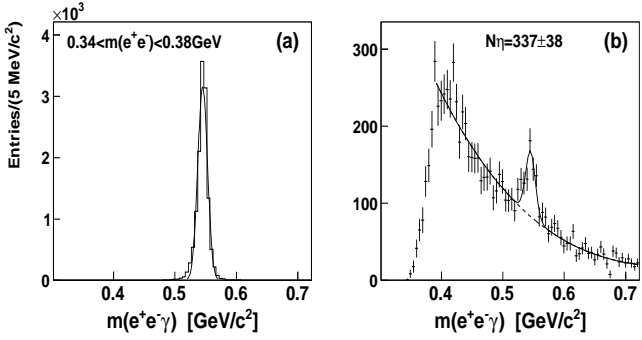}
\caption{
 Same as Fig.~\protect\ref{fig:eegz34_eta_cth_2007_fit0_m340_380}, but after
 applying the tighter cuts on $N$ and $R$
 (depicted by blue solid lines in Fig.~\protect\ref{fig:omega_pi0ee_pi_cuts}).
}
 \label{fig:eegz34_eta_cth_2007_fit2_m340_380} 
\end{figure}

 The fitting procedure for $\eta \to e^+e^-\gamma$ and the impact of selection
 criteria on the background is illustrated in
 Figs.~\ref{fig:eegz34_eta_cth_2007_fit10_m30_100}--\ref{fig:eegz34_eta_cth_2007_fit1_4_m30_100}
 and Figs.~\ref{fig:eegz34_eta_cth_2007_fit0_m340_380}--\ref{fig:eegz34_eta_cth_2007_fit2_m340_380}.
 Figure~\ref{fig:eegz34_eta_cth_2007_fit10_m30_100} shows all
 $\gamma p\to e^+e^-\gamma p$ candidates in the $m(e^+e^-)$
 range from 30 to 100~MeV/$c^2$. These were initially selected with
 the kinematic-fit CL only, also allowing both $e^+$ and $e^-$ to be identified
 with the same PID element.
 Figure~\ref{fig:eegz34_eta_cth_2007_fit10_m30_100}(a) depicts the $m(e^+e^-\gamma)$
 invariant-mass distribution for the MC simulation of
 $\gamma p\to \eta p \to e^+e^-\gamma p$ fitted with a Gaussian.
 The experimental distribution after subtracting both the random background
 and the background remaining from $\gamma p\to \eta p \to \gamma\gamma p$
 is shown by crosses in Fig.~\ref{fig:eegz34_eta_cth_2007_fit10_m30_100}(b).
 The distribution for the $\eta \to \gamma\gamma$ background is normalized
 to the number of subtracted events and is shown in the same figure by
 a red solid line.
 The subtraction normalization was based on the number of events generated
 for $\gamma p\to \eta p \to \gamma\gamma p$ and the number
 of $\gamma p\to \eta p$ events produced in the same experiment.
 The experimental distribution was fitted with the sum
 of a Gaussian for the $\eta \to e^+e^-\gamma$ peak and
 a polynomial of order four for the background.
 In this fit, the centroid and width of the Gaussian were fixed
 to the values obtained from the previous Gaussian fit
 to the $\gamma p\to \eta p \to e^+e^-\gamma p$ MC simulation,
 which is shown in Fig.~\ref{fig:eegz34_eta_cth_2007_fit10_m30_100}(a).
 As seen, the Gaussian parameters obtained from fitting
 to the MC simulation suit the experimental peak well.
 This confirms the agreement of the experimental data and the MC
 simulation in the energy calibration of the calorimeters
 and their resolution.
 The order of the polynomial was chosen to be sufficient for a reasonable
 description of the background distribution in the range of fitting.

 The number of $\eta \to e^+e^-\gamma$ decays
 in the experimental $m(e^+e^-\gamma)$ spectra
 was determined from the area under the Gaussian. For consistency,
 the $\gamma p\to \eta p \to e^+e^-\gamma p$ detection efficiency
 in each $m(e^+e^-)$ bin was obtained in the same way, i.e.,
 based on the $m(e^+e^-\gamma)$ spectrum for the MC simulation fitted
 with a Gaussian, instead of using the number of entries in
 this spectrum. For the selection criteria and the $m(e^+e^-)$ range used
 to obtain the spectra shown in Fig.~\ref{fig:eegz34_eta_cth_2007_fit10_m30_100},
 the averaged detection efficiency determined for
 $\gamma p\to \eta p \to e^+e^-\gamma p$ in this manner is 33.1\%.

 Figure~\ref{fig:eegz34_eta_cth_2007_fit0_m30_100} illustrates
 the effect of requiring both $e^+$ and $e^-$ to be identified
 by different PID elements.
 As seen, compared to Fig.~\ref{fig:eegz34_eta_cth_2007_fit10_m30_100}(b),
 the $\eta \to \gamma\gamma$ background becomes very small.
 The signal-to-background ratio improves significantly as well,
 whereas the $\gamma p\to \eta p \to e^+e^-\gamma p$ detection efficiency
 decreases to 25.8\%. The results for the $\eta \to e^+e^-\gamma$ yield obtained
 with and without adding events with $e^+$ and $e^-$ identified by
 the same PID element showed good agreement within the fit uncertainties,
 confirming the reliability of the $\eta \to \gamma\gamma$ background subtraction.

 The almost full elimination of the background contributions under the
 $\eta \to e^+e^-\gamma$ peak in this $m(e^+e^-)$ range can be obtained
 by applying the $dE/dx$ PID cut selecting only events with the elements
 having a deposit corresponding to a single $e^{+/-}$, and also requiring
 the final-state proton to be detected.
 The spectra obtained with such cuts are shown
 in Fig.~\ref{fig:eegz34_eta_cth_2007_fit1_4_m30_100}.
 Although the $\gamma p\to \eta p \to e^+e^-\gamma p$ detection efficiency
 decreases to 22.3\%, the smallness of the background under
 the $\eta \to e^+e^-\gamma$ peak makes it possible to measure
 the $\eta\to \gamma\gamma^* \to \gamma e^+e^-$ angular dependence
 of the virtual photon decaying into a lepton pair and to compare it
 with Eq.~(\ref{eqn:dtheta}).  

 The experimental results for such an angular dependence are
 illustrated in Fig.~\ref{fig:eta_eeg_cth2e_a2_2007_3x1}.
 Figure~\ref{fig:eta_eeg_cth2e_a2_2007_3x1}(a) shows
 the experimental $\cos\theta^*$ distribution obtained for
 the events in the $\eta \to e^+e^-\gamma$ peak from
 Fig.~\ref{fig:eegz34_eta_cth_2007_fit1_4_m30_100}(b).
 The corresponding angular acceptance determined from
 the MC simulation is depicted in
 Fig.~\ref{fig:eta_eeg_cth2e_a2_2007_3x1}(b).
 The experimental $\cos\theta^*$ distribution corrected
 for the acceptance is depicted in
 Fig.~\ref{fig:eta_eeg_cth2e_a2_2007_3x1}(c), showing reasonable agreement
 with the expected $1 + \cos^2\theta^*$ dependence.
 Because $e^+$ and $e^-$ cannot be separated in the present experiment,
 the angles of both leptons were used to measure the dilepton decay
 dependence, which resulted in a symmetric shape with respect to
 $\cos\theta^*=0.$
\begin{figure}
\includegraphics[width=0.45\textwidth,height=4.35cm,bbllx=1.cm,bblly=.5cm,bburx=19.5cm,bbury=9.cm]{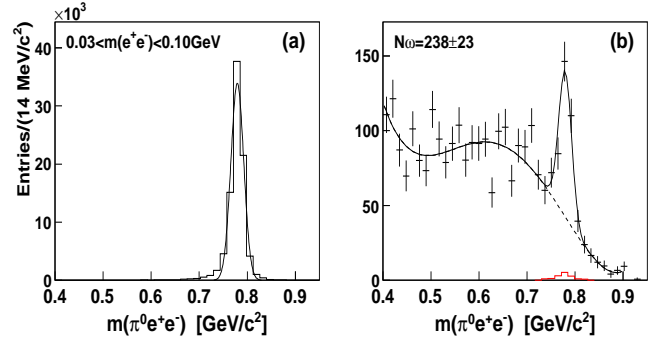}
\caption{ (Color online)
 $m(\pi^0e^+e^-)$ invariant-mass distributions obtained for the $m(e^+e^-)$
 range from 30 to 100~MeV/$c^2$ with $\gamma p\to \pi^0e^+e^-p$ candidates
 initially selected with the kinematic fit
 (a)~MC simulation of $\gamma p\to \omega p \to \pi^0e^+e^-p$ fitted with
     a Gaussian;
 (b)~experimental events (Run-I) after subtracting the random background
     and the remaining background from $\omega \to \pi^0\gamma$ decays
     (with both the $\pi^0\to\gamma\gamma$ and $\pi^0\to e^+e^-\gamma$ decay modes).
     The distribution for the $\omega \to \pi^0\gamma$ background, shown
     by a red solid line, is normalized to the number of subtracted events.
     The experimental distribution is fitted with the sum of a Gaussian
     for the $\omega\to\pi^0e^+e^-$ peak and a polynomial of order five
     for the background. The total fit is depicted by a solid line,
     and the dashed line shows the background under the peak. 
}
 \label{fig:pi0eez5_omega_cth_2007_fit0_1_m30_100} 
\end{figure}
\begin{figure}
\includegraphics[width=0.45\textwidth,height=4.35cm,bbllx=1.cm,bblly=.5cm,bburx=19.5cm,bbury=9.cm]{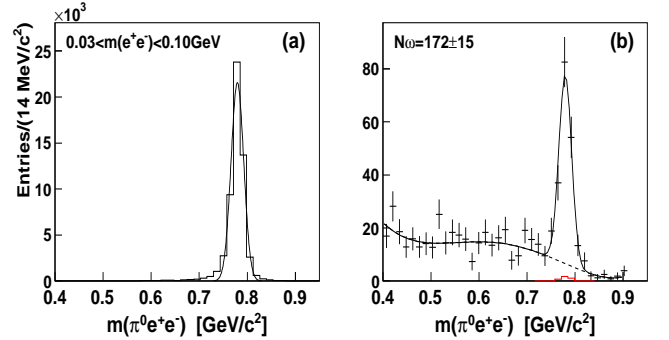}
\caption{ (Color online)
 Same as Fig.~\protect\ref{fig:pi0eez5_omega_cth_2007_fit0_1_m30_100}, but after
 applying the softer cuts on $N$ and $R$
 (depicted by red dashed lines in Fig.~\protect\ref{fig:omega_pi0ee_pi_cuts})
 and the $dE/dx$ cut rejecting events with twice the magnitude
 of a signal from a single $e^{+/-}$ in one PID element.
}
 \label{fig:pi0eez5_omega_cth_2007_fit6_101_m30_100} 
\end{figure}

 At higher $m(e^+e^-)$ masses, in addition to $\gamma p\to \pi^0\pi^0p$
 events, there is background from $\eta \to \pi^+\pi^-\gamma$ and 
 $\eta \to \pi^+\pi^-\pi^0$ decays. These decays do not mimic
 the $\eta \to e^+e^-\gamma$ peak, but, without suppression of
 the $\pi^{+/-}$ background, the signal becomes
 comparable with the statistical fluctuations of the background events.
 The suppression of this background with the cuts on the multiplicity
 $N$ of the crystals forming $e^{+/-}$ clusters and their effective
 radius $R$ is illustrated in 
 Figs.~\ref{fig:eegz34_eta_cth_2007_fit0_m340_380}--\ref{fig:eegz34_eta_cth_2007_fit2_m340_380}.
\begin{figure*}
\includegraphics[width=0.85\textwidth]{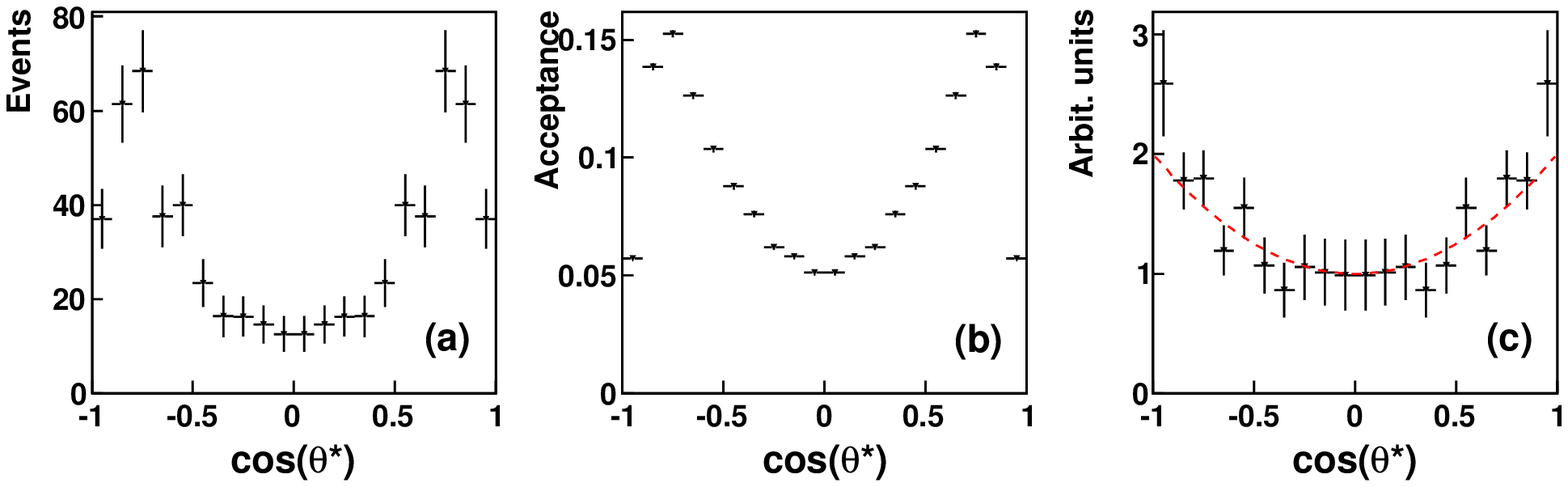}
\caption{ (Color online)
 The $\omega\to \pi^0\gamma^* \to \pi^0 e^+e^-$ angular dependence
 (in the $\omega$ rest frame)
 of the virtual photon decaying into a lepton pair, with $\theta^*$
 being the angle between the direction of one of the leptons in
 the virtual-photon (or the dilepton) rest frame and the direction
 of the dilepton system (which is opposite to the $\pi^0$ direction):
 (a) experimental events from the $\omega \to \pi^0e^+e^-$ peak;
 (b) angular acceptance based on the MC simulation;
 (c) the experimental spectrum corrected for the acceptance
     and normalized for comparing to the $1 + \cos^2\theta^*$ dependence
     (shown by a red dashed line).
 Because $e^+$ and $e^-$ cannot be separated in the present experiment,
 the angles of both leptons were used, resulting in a symmetric shape
 with respect to $\cos\theta^*=0.$
}
 \label{fig:omega_pi0ee_cthe_a2_2007_3x1} 
\end{figure*}
 Figure~\ref{fig:eegz34_eta_cth_2007_fit0_m340_380} shows
 $\gamma p\to e^+e^-\gamma p$ candidates selected with
 the kinematic-fit CL in the $m(e^+e^-)$ region, where the magnitude of
 the $\eta \to e^+e^-\gamma$ peak is still sufficient to see it above
 a large background. The result of applying the softer cuts
 on $N$ and $R$ (depicted by red dashed lines in Fig.~\ref{fig:omega_pi0ee_pi_cuts})
 is demonstrated in Fig.~\ref{fig:eegz34_eta_cth_2007_fit1_m340_380}, showing
 a significant improvement in the signal-to-background ratio.
 The further improvement with the tighter cuts on $N$ and $R$
 (depicted by blue solid lines in Fig.~\ref{fig:omega_pi0ee_pi_cuts})
 is demonstrated in Fig.~\ref{fig:eegz34_eta_cth_2007_fit2_m340_380}.
 The fits with the suppressed background in this $m(e^+e^-)$ range are
 more reliable, even if the $\gamma p\to \eta p \to e^+e^-\gamma p$
 detection efficiency decreases from 33.4\% to 27.8\% after applying
 the softer cuts, and to 24.0\% after applying the tighter cuts.
 It was checked that the results for
 the $\eta \to e^+e^-\gamma$ yield obtained with and without
 cuts on $N$ and $R$ were in good agreement within the fit uncertainties,
 confirming the reliability of the method based on the difference in
 the features of $e^{+/-}$ and $\pi^{+/-}$ clusters.
 Note that the $\eta \to \gamma\gamma$ background is negligibly small
 in this range of $m(e^+e^-)$ masses, even with both $e^+$ and $e^-$ being
 identified by the same PID elements.

\begin{figure}
\includegraphics[width=0.45\textwidth,height=4.35cm,bbllx=1.cm,bblly=.5cm,bburx=19.5cm,bbury=9.cm]{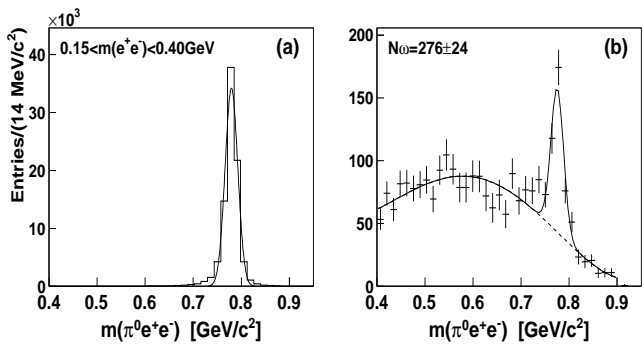}
\caption{
 Same as Fig.~\protect\ref{fig:pi0eez5_omega_cth_2007_fit0_1_m30_100}, but for 
 the $m(e^+e^-)$ range from 150 to 400~MeV/$c^2$ and the tighter cuts on $N$ and $R$
 (depicted by blue solid lines in Fig.~\protect\ref{fig:omega_pi0ee_pi_cuts}).
 The $\omega \to \pi^0\gamma$ background is not shown in (b) because it is negligibly small.
}
 \label{fig:pi0eez5_omega_cth_2007_fit0_m150_400} 
\end{figure}

\begin{figure}
\includegraphics[width=0.45\textwidth,height=4.35cm,bbllx=1.cm,bblly=.5cm,bburx=19.5cm,bbury=9.cm]{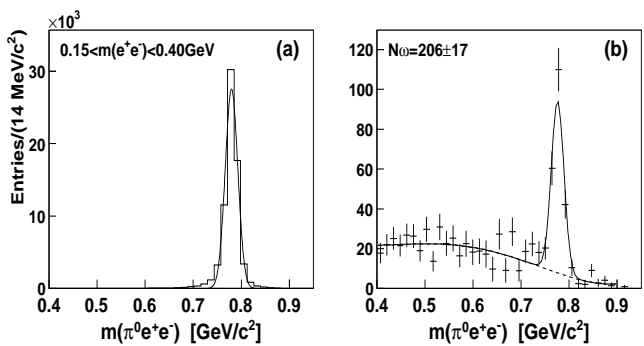}
\caption{
 Same as Fig.~\protect\ref{fig:pi0eez5_omega_cth_2007_fit0_m150_400}, but after
 applying the $dE/dx$ cut rejecting events with twice the magnitude
 of a signal from a single $e^{+/-}$ in one PID element.
}
 \label{fig:pi0eez5_omega_cth_2007_fit6_m150_400} 
\end{figure}

 Because the $\omega\to\pi^0e^+e^-$ decays were analyzed in the same data
 sets and by using the same cuts, the systematic uncertainties caused
 by these cuts should be the same as for $\eta \to e^+e^-\gamma$.
 Additional tests were made for $m(e^+e^-)$ ranges with
 less background and wide $m(e^+e^-)$ bins, giving smaller
 statistical uncertainties in the results. The fitting procedure
 for $\omega\to\pi^0e^+e^-$ (which is very similar to $\eta \to e^+e^-\gamma$)
 and some of the tests, including the $\omega\to \pi^0\gamma^* \to \pi^0 e^+e^-$
 angular dependence of the virtual photon decaying into a lepton pair, 
 are illustrated in
 Figs.~\ref{fig:pi0eez5_omega_cth_2007_fit0_1_m30_100}--\ref{fig:pi0eez5_omega_cth_2007_fit6_m150_400}.

 Figure~\ref{fig:pi0eez5_omega_cth_2007_fit0_1_m30_100} shows
 $\gamma p\to \pi^0e^+e^- p$ candidates selected only with
 the kinematic-fit CL,
 for the $m(e^+e^-)$ region below the $\pi^0$ mass,
 avoiding very large background from $\gamma p\to \pi^0\pi^0p$.
 Figure~\ref{fig:pi0eez5_omega_cth_2007_fit0_1_m30_100}(a) depicts the $m(\pi^0e^+e^-)$
 invariant-mass distribution for the MC simulation of
 $\gamma p\to \omega p \to \pi^0e^+e^-p$ fitted with a Gaussian.
 The choice of the normal distribution for fitting the $\omega$ peak
 is motivated by the facts that the BW shape of the $\omega$ signal
 is severely cut by phase space near threshold
 and the $m(\pi^0e^+e^-)$ resolution strongly dominates
 the $\omega$-meson width ($\Gamma=8.49$~MeV~\cite{PDG}).
 A similar approach was successfully used for fitting the $\omega\to \pi^0\gamma$
 peak above background while measuring $\omega$ photoproduction with the same data
 set~\cite{a2_omegap_2015}.
 The experimental distribution after subtracting both the random background
 and the background remaining from $\omega \to \pi^0\gamma$ decays
 (with both the $\pi^0\to\gamma\gamma$ and $\pi^0\to e^+e^-\gamma$ decay modes)
 is shown by crosses in Fig.~\ref{fig:pi0eez5_omega_cth_2007_fit0_1_m30_100}(b).
 The distribution for the $\omega \to \pi^0\gamma$ background is normalized
 to the number of subtracted events and is shown in the same figure by
 a red solid line.
 The subtraction normalization was based on the number of events generated
 for $\gamma p\to \omega p \to \pi^0\gamma p$ and the number
 of $\gamma p\to \omega p$ events produced in the experiment.
 The experimental distribution was fitted with the sum
 of a Gaussian for the $\omega \to \pi^0e^+e^-$ peak and
 a polynomial of order five for the background.
 In this fit, the centroid and width of the Gaussian were fixed
 to the values obtained from the previous Gaussian fit
 to the $\gamma p\to \omega p \to \pi^0e^+e^- p$ MC simulation,
 which is shown in Fig.~\ref{fig:pi0eez5_omega_cth_2007_fit0_1_m30_100}(a).
 Similar to $\eta \to e^+e^-\gamma$, the number of $\omega \to \pi^0e^+e^-$
 decays in the MC and experimental $m(\pi^0e^+e^-)$ spectra
 was determined from the area under the Gaussian.
 For the selection criteria used to obtain
 the spectra in Fig.~\ref{fig:pi0eez5_omega_cth_2007_fit0_1_m30_100}
 and the given $m(e^+e^-)$ range, the averaged detection efficiency
 determined for $\omega \to \pi^0e^+e^-$ is 14.5\%.

 Figure~\ref{fig:pi0eez5_omega_cth_2007_fit6_101_m30_100} illustrates
 the impact of the PID $dE/dx$ cut and the softer cuts on $N$ and $R$
 on suppressing background under the $\omega$ peak.
 As seen, compared to Fig.~\ref{fig:pi0eez5_omega_cth_2007_fit0_1_m30_100}(b), 
 the quantity of background events becomes smaller by a factor of 5
 (resulting in a more reliable fit to the signal peak), whereas the detection efficiency
 for $\omega \to \pi^0e^+e^-$ decreases to 9.1\%.

 Although the level of the background remaining under the $\omega \to \pi^0e^+e^-$
 peak is not negligibly small, it is still possible to check
 the $\omega\to \pi^0\gamma^* \to \pi^0 e^+e^-$ angular dependence
 of the virtual photon decaying into a lepton pair, compared to Eq.~(\ref{eqn:dtheta}).  
 The experimental results for such an angular dependence are
 illustrated in Fig.~\ref{fig:omega_pi0ee_cthe_a2_2007_3x1}.

 Figure~\ref{fig:omega_pi0ee_cthe_a2_2007_3x1}(a) shows
 the experimental $\cos\theta^*$ distribution obtained for
 the events in the $\omega \to \pi^0e^+e^-$ peak from
 Fig.~\ref{fig:pi0eez5_omega_cth_2007_fit6_101_m30_100}(b).
 The corresponding angular acceptance determined from
 the MC simulation is depicted in
 Fig.~\ref{fig:omega_pi0ee_cthe_a2_2007_3x1}(b).
 The experimental $\cos\theta^*$ distribution corrected
 for the acceptance is depicted in
 Fig.~\ref{fig:omega_pi0ee_cthe_a2_2007_3x1}(c), showing,
 for the very limited statistics and the remaining background, reasonable
 agreement with the expected $1 + \cos^2\theta^*$ dependence.

 Figures~\ref{fig:pi0eez5_omega_cth_2007_fit0_m150_400}
 and~\ref{fig:pi0eez5_omega_cth_2007_fit6_m150_400} illustrate fits
 for the $m(e^+e^-)$ region above the $\pi^0$ mass, with
 $\gamma p\to \pi^0e^+e^- p$ candidates selected after
 the tighter cuts on $N$ and $R$ (better suppressing
 the $\omega\to \pi^0\pi^+\pi^-$ background), and without and
 with $dE/dx$ PID cuts.
 As seen from these figures, the quantity of background events
 in the vicinity of the $\omega$ peak becomes smaller by a factor
 of more than 4 after applying the $dE/dx$ PID cut, whereas the detection
 efficiency, averaged in this $m(e^+e^-)$ range, decreases from 18.9\% to 15.1\%.
 The $\omega \to \pi^0\gamma$ background is negligibly small
 in this range of $m(e^+e^-)$ masses, and it is not shown in these figures.
 The fits made without and with $dE/dx$ PID cuts showed good agreement within
 the fit uncertainties, confirming the reliability of tests made
 with the $\eta \to e^+e^-\gamma$ decay.

 To measure the $\eta \to e^+e^-\gamma$ and $\omega \to \pi^0e^+e^-$ yields
 as a function of the invariant mass $m(e^+e^-)$, the corresponding
 candidate events were divided into several $m(e^+e^-)$ bins,
 separately for Run-I and Run-II.
 The available statistics and the level of background for
 $\eta \to e^+e^-\gamma$ decays enabled
 division of the $m(e^+e^-)$ range from 30 to 490~MeV/$c^2$ into 34 bins,
 with bin widths increasing from 10 MeV/$c^2$ at the lowest masses
 to 30 MeV/$c^2$ at the highest masses.   
 To measure the $\omega \to \pi^0e^+e^-$ decay, the $m(e^+e^-)$ range
 from 30 to 630~MeV/$c^2$ was divided into 14 bins,
 with bin widths increasing from 20 MeV/$c^2$ at the lowest masses
 to 60 MeV/$c^2$ at the highest masses. The size and width of
 the $m(e^+e^-)$ bins for Run-I and Run-II were identical,
 which later allowed the results from both runs to be combined.    
 The fitting procedure was the same as those used to check
 the systematic uncertainties caused by various selection criteria. 

\section{Results and discussion}
  \label{sec:Results}

\subsection{TFF results and their uncertainties}
\label{subsec:Results-I}
\begin{figure*}
\includegraphics[width=0.925\textwidth]{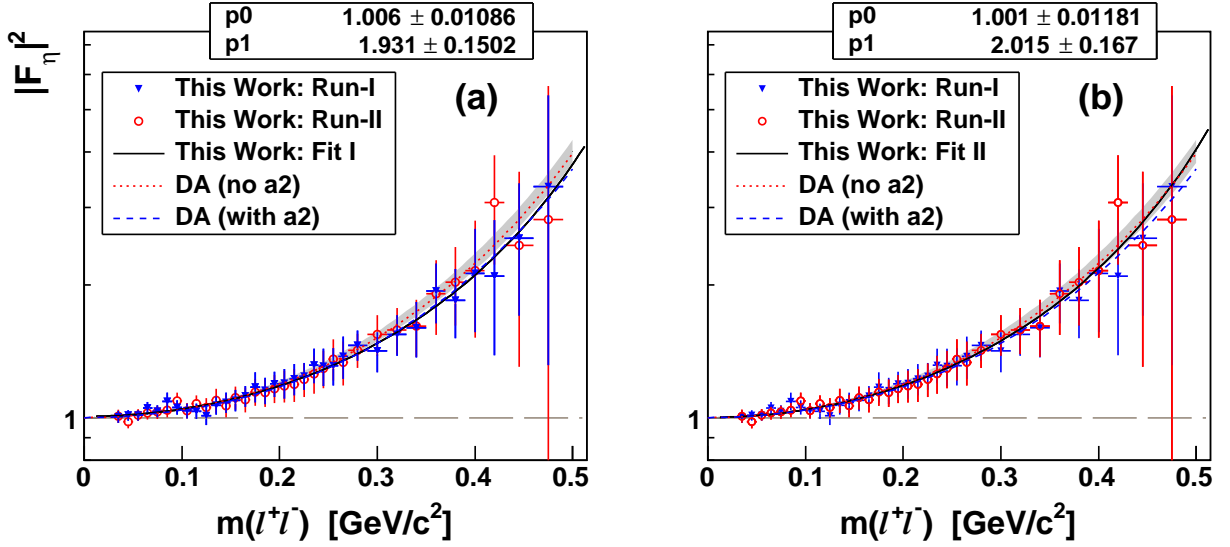}
\caption{ (Color online)
  Comparison of the $|F_{\eta}(m_{\ell^+\ell^-})|^2$ results
 obtained individually from the analyses of Run-I (blue filled triangles)
 and Run-II  (red open circles) with each other and with the two
 solutions for the DA calculations by the J\"ulich
 group~\protect\cite{Xiao_2015,Hanhart}.
 The solution without including the $a_2$-meson contribution
 is shown by a red dotted line with an error band, and the solution
 involving the $a_2$ contribution is shown by a blue dashed line.
 The pole-approximation fits (black solid lines) to the results
 of Run-I and Run-II are depicted in panels (a) and (b), respectively.
 The fit parameter $p0$ reflects the general normalization of the data points,
 and $p1$ is the slope parameter $\Lambda^{-2}$~[GeV$^{-2}$].
 For a better comparison of the magnitudes of total uncertainties from the
 two data sets, the error bars of Run-I are plotted in (a) on the top of
 the error bars of Run-II, and the other way around in (b).
}
 \label{fig:eta_tff_cth_a2_07_vs_09_ebst_sys2_2x1} 
\end{figure*}
 The total number of $\eta \to e^+e^-\gamma$ and $\omega \to \pi^0e^+e^-$
 decays initially produced in each $m(e^+e^-)$ bin was obtained
 by correcting the number of decays observed in each bin
 with the corresponding detection efficiency.
 Values of $d\Gamma(\eta \to e^+e^-\gamma)/dm(e^+e^-)$
 and $d\Gamma(\omega \to \pi^0e^+e^-)/dm(e^+e^-)$ for every fit
 were obtained from those initial numbers of decays by taking into
 account the full decay width of $\eta$ and $\omega$~\cite{PDG},
 the total number of $\eta$ and $\omega$ mesons produced in the same data
 sets~\cite{etamamic,a2_omegap_2015}, and the width of the corresponding
 $m(e^+e^-)$ bin. The uncertainty in an individual $d\Gamma/dm(e^+e^-)$
 value from a particular fit was based on the uncertainty in the number
 of decays determined by this fit (i.e., the uncertainty in the area
 under the Gaussian).
 The systematic uncertainties in the $d\Gamma/dm(e^+e^-)$ value were estimated
 for each individual $m(e^+e^-)$ bin by repeating its fitting procedure
 several times after refilling the $m(e^+e^-\gamma)$ spectra with different
 combinations of selection criteria, which were used to improve
 the signal-to-background ratio, or after slight changes in the parametrization
 of the background under the signal peak. The changes in selection criteria
 included cuts on the kinematic-fit CL (such as 2\%, 5\%, 10\%, 15\%, and 20\%),
 different cuts on PID $dE/dx$, $N$, $R$, and $z$.
 As in Ref.~\cite{eta_tff_a2_2014}, the $\eta \to e^+e^-\gamma$ results
 were also checked for excluding three-cluster events (no final-state proton
 detected) from the analysis. The requirement of 
 making several fits for each $m(e^+e^-)$ bin provided a check on
 the stability of the $d\Gamma/dm(e^+e^-)$ results.
 The average of the results of all fits made for one bin was then used
 to obtain final $d\Gamma/dm$ values that were more reliable than the results
 based on just one so-called best fit, which was made with a combination
 of selection criteria, giving the optimal number
 of events in the signal peak with respect to the background level under it.
 Typically, such a best fit gives the largest ratio between
 the corresponding $d\Gamma/dm$ value and its uncertainty.
\begin{table*}
\caption
[tab:etatff]{
 Results of this work for the $\eta$ TFF, $|F_{\eta}|^2$, as a function of
 the invariant mass $m(e^+e^-)$.
 } \label{tab:etatff}
\begin{ruledtabular}
\begin{tabular}{|c|c|c|c|c|c|c|c|} 
\hline
 $m(e^+e^-)$~[MeV/$c^2$]
 & $35\pm5$ & $45\pm5$ & $55\pm5$ & $65\pm5$ & $75\pm5$ & $85\pm5$ & $95\pm5$ \\
\hline
 $|F_{\eta}|^2$ 
 & $1.006\pm0.024$ & $0.999\pm0.022$ & $1.013\pm0.021$ & $1.037\pm0.024$
 & $1.032\pm0.024$ & $1.057\pm0.031$ & $1.070\pm0.030$ \\
\hline
\hline
 $m(e^+e^-)$~[MeV/$c^2$]
 & $105\pm5$ & $115\pm5$ & $125\pm5$ & $135\pm5$ & $145\pm5$ & $155\pm5$ & $165\pm5$ \\
\hline
 $|F_{\eta}|^2$ 
 & $1.038\pm0.029$ & $1.052\pm0.032$ & $1.030\pm0.035$ & $1.077\pm0.041$
 & $1.074\pm0.042$ & $1.101\pm0.045$ & $1.111\pm0.046$ \\
\hline
\hline
 $m(e^+e^-)$~[MeV/$c^2$]
  & $175\pm5$ & $185\pm5$ & $195\pm5$ & $205\pm5$  & $215\pm5$ & $225\pm5$ & $235\pm5$ \\
\hline
 $|F_{\eta}|^2$ 
 & $1.157\pm0.060$ & $1.146\pm0.057$ & $1.179\pm0.057$ & $1.189\pm0.067$
 & $1.207\pm0.072$ & $1.234\pm0.067$ & $1.288\pm0.085$ \\
\hline
\hline
 $m(e^+e^-)$~[MeV/$c^2$]
 & $245\pm5$ & $255\pm5$ & $265\pm5$ & $280\pm10$ & $300\pm10$ & $320\pm10$ & $340\pm10$ \\
\hline
 $|F_{\eta}|^2$ 
 & $1.300\pm0.090$ & $1.331\pm0.095$ & $1.357\pm0.107$ & $1.443\pm0.085$
 & $1.473\pm0.110$ & $1.561\pm0.124$ & $1.607\pm0.166$ \\
\hline
\hline
 $m(e^+e^-)$~[MeV/$c^2$]
 & $360\pm10$ & $380\pm10$ & $400\pm10$ & $420\pm10$ & $445\pm15$ & $475\pm15$ & \\
\hline
 $|F_{\eta}|^2$ 
 & $1.925\pm0.232$ & $1.916\pm0.257$ & $2.137\pm0.421$ & $2.495\pm0.547$
 & $2.519\pm0.685$ & $3.17\pm1.65$ & \\
\hline
\end{tabular}
\end{ruledtabular}
\end{table*}

 Because the fits for a given $m(e^+e^-)$ bin with different selection criteria
 or different background parametrizations were based on the same initial data
 sample, the corresponding $d\Gamma/dm$ results were correlated and could not be
 considered as independent measurements for calculating the uncertainty
 in the averaged $d\Gamma/dm$ value. Thus, this uncertainty was taken from the best
 fit for the given $m(e^+e^-)$ bin, which was a conservative estimate of the uncertainty
 in the averaged $d\Gamma/dm$ value.
 The systematic uncertainty in this $d\Gamma/dm$ value
 was taken as the root mean square of the results from all fits
 made for this bin. The total uncertainty in this $d\Gamma/dm$ value
 was calculated by adding in quadrature its fit (partially reflecting
 experimental statistics in the bin) and systematic uncertainties.
 The overall statistics of $5.4\cdot 10^4$
 $\eta \to e^+e^-\gamma$ decays involved in all the fits provided
 quite small fit uncertainties, with the average magnitude
 of the systematic uncertainties being $\sim$35\% of the fit uncertainties.
 Because the overall statistics for $\omega \to \pi^0e^+e^-$ were only
 $1.1\cdot 10^3$ decays, the total uncertainties were dominated by
 the fit uncertainties, with average magnitude of
 the systematic uncertainties being $\sim$20\% of the fit uncertainties.
 In the end, the $d\Gamma/dm(e^+e^-)$ results from Run-I and Run-II,
 which were independent measurements, were combined as a weighted average
 with weights taken as inverse values of their total uncertainties in quadrature.

 The results for $|F_{\eta}(m_{e^+e^-})|^2$ and $|F_{\omega\pi^0}(m_{e^+e^-})|^2$ were obtained
 by dividing the combined results for $d\Gamma(\eta \to e^+e^-\gamma)/dm(e^+e^-)$
 and $d\Gamma(\omega \to \pi^0e^+e^-)/dm(e^+e^-)$ by the corresponding QED terms
 from Eqs.~(\ref{eqn:dgdm_eta}) and~(\ref{eqn:dgdm_omega}), and using
 the $\eta\to\gamma\gamma$ and $\omega\to\pi^0\gamma$ branching ratios
 from RPP~\cite{PDG}.
 To check the consistency of the individual TFF results obtained from Run-I and
 Run-II, the corresponding $d\Gamma/dm(e^+e^-)$ results were recalculated
 into $|F_{\eta}(m_{e^+e^-})|^2$ and $|F_{\omega\pi^0}(m_{e^+e^-})|^2$ as well.

\subsection{comparison of $\eta$ results with other data and calculations}
\label{subsec:Results-II}
 The individual $|F_{\eta}(m_{e^+e^-})|^2$ results from Run-I and Run-II
 are compared in Fig.~\ref{fig:eta_tff_cth_a2_07_vs_09_ebst_sys2_2x1}.
 For a better comparison of the magnitudes
 of total uncertainties in both the measurements, with the same $m(e^+e^-)$
 binning, the experimental results are plotted twice.
 In Fig.~\ref{fig:eta_tff_cth_a2_07_vs_09_ebst_sys2_2x1}(a), the error bars of
 Run-I are plotted on the top of the error bars of Run-II, and the other
 way around in Fig.~\ref{fig:eta_tff_cth_a2_07_vs_09_ebst_sys2_2x1}(b).
 Correspondingly, the fit to the $|F_{\eta}|^2$ results of Run-I with
 Eq.~(\protect\ref{eqn:Fm}) is shown in Fig.~\ref{fig:eta_tff_cth_a2_07_vs_09_ebst_sys2_2x1}(a),
 and of Run-II in Fig.~\ref{fig:eta_tff_cth_a2_07_vs_09_ebst_sys2_2x1}(b).
 The fits are made with two free parameters, one of which, $p1$,
 is $\Lambda^{-2}$, and the other, $p0$, reflects
 the general normalization of the data points.
 For example, the latter parameter could be different from $p0=1$
 because of the uncertainty in the determination of
 the experimental number of $\eta$ mesons produced.
 Another possible reason for $p0$ to be slightly more than one
 is radiative corrections for the QED differential decay
 rate at low $q$, the magnitude of which is expected to be $\sim$1\%.
\begin{figure*}
\includegraphics[width=0.925\textwidth]{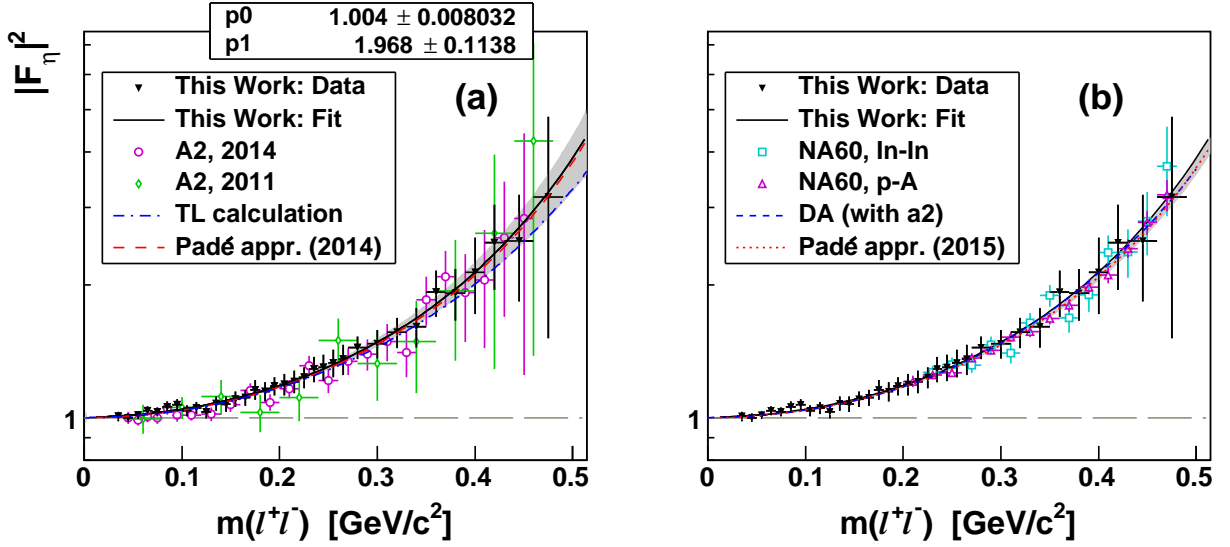}
\caption{ (Color online)
  $|F_{\eta}(m_{\ell^+\ell^-})|^2$ results (black filled triangles)
  combined from Run-I and Run-II and their pole-approximation fit
 (black solid line, with $p0$ and $p1$ being the normalization
 and the slope parameter $\Lambda^{-2}$, respectively)
 are compared to previous measurements and various
 theoretical calculations.  The former results by the A2 Collaboration from
  Ref.~\protect\cite{eta_tff_a2_2014} (open magenta circles)
  and Ref.~\protect\cite{eta_tff_a2_2011} (open green diamonds)
  are shown in panel (a). The results of NA60 obtained in
  peripheral In--In data~\protect\cite{NA60_2009} and in
  p--A collisions~\protect\cite{NA60_2016} are shown in panel (b).
  The calculation from Ref.~\protect\cite{Ter10} is shown in (a) by
  a blue dash-dotted line.
  The most recent DA calculation by the J\"ulich group~\protect\cite{Xiao_2015}
  is shown in (b) by a blue dashed line.
  The calculations by the Mainz group with the Pad\'e approximants
  are shown in (a) for their previous solution~\protect\cite{Esc13}
 (red dashed line with an error band) and in (b) for their latest
  solution~\protect\cite{Escribano_2015} (red dotted line with an error band).
}
 \label{fig:eta_tff_cth_a2_ebst_syst_exp_th2_2x1} 
\end{figure*}

 The correlation between the two parameters results
 in a larger fit error for $\Lambda^{-2}$.
 However, this error then includes the systematic uncertainty
 in the general normalization of the data points.
 Because all $|F_{\eta}|^2$ results are obtained with their total
 uncertainties, the fit error for $\Lambda^{-2}_{\eta}$ gives its
 total uncertainty as well.
 As seen in Fig.~\ref{fig:eta_tff_cth_a2_07_vs_09_ebst_sys2_2x1},
 the fits to both Run-I and Run-II results give normalization
 parameters compatible with the expected values, indicating
 the good quality of the results. 
 A value of the second parameter obtained for Run-I,
 $p1=(1.93\pm0.15_{\mathrm{tot}})$~GeV$^{-2}$, is slightly smaller than 
 the value from Ref.~\cite{eta_tff_a2_2014},
 $\Lambda^{-2}_{\eta}=(1.95\pm 0.15_{\mathrm{stat}}\pm 0.10_{\mathrm{syst}})
 =(1.95\pm 0.18_{\mathrm{tot}})$~GeV$^{-2}$, also obtained from the analysis
 of Run-I, but is in good agreement within the uncertainties,
 the magnitude of which became somewhat smaller as well.
 The value, $p1=(2.02\pm0.17_{\mathrm{tot}})$~GeV$^{-2}$, obtained
 for Run-II is slightly larger than both the present and the previous
 results from Run-I, but is in good agreement
 within the uncertainties. The magnitude of the difference
 in the $\Lambda^{-2}_{\eta}$ results obtained for Run-I and Run-II
 is comparable to the uncertainties in the theoretical predictions
 for $\Lambda^{-2}_{\eta}$. As an example, the most recent
 calculations with the dispersive analysis (DA) by the J\"ulich group
 are shown in Fig.~\ref{fig:eta_tff_cth_a2_07_vs_09_ebst_sys2_2x1} for their
 new solution~\cite{Xiao_2015}, obtained after including the $a_2$-meson
 contribution in the analysis, and their previous solution without
 it~\cite{Hanhart}.  As seen, the fit of Run-II practically overlaps with
 the calculation without $a_2$, and the fit of Run-I
 is very close to the calculation involving the $a_2$ contribution.

 The $|F_{\eta}(m_{e^+e^-})|^2$ results combined from Run-I and Run-II
 are compared to previous measurements and various theoretical calculations
 in Fig.~\ref{fig:eta_tff_cth_a2_ebst_syst_exp_th2_2x1}.
 The numerical values for the combined $|F_{\eta}(m_{e^+e^-})|^2$ results
 are listed in Table~\ref{tab:etatff}.
 As seen in Fig.~\ref{fig:eta_tff_cth_a2_ebst_syst_exp_th2_2x1},
 the present $|F_{\eta}(m_{e^+e^-})|^2$ results are in good agreement, within
 the error bars, with all previous measurements based on
 $\eta\to e^+e^-\gamma$ and $\eta\to\mu^+\mu^-\gamma$ decays.
 The pole-approximation fit to the present $|F_{\eta}|^2$ data points yields
\begin{equation}
 \Lambda^{-2}_{\eta}=(1.97\pm 0.11_{\mathrm{tot}}) ~\mathrm{GeV}^{-2},
\label{eqn:Lam2_eta_this_work}
\end{equation}
 which is also in very good agreement within the uncertainties with the
 results reported in Refs.~\cite{NA60_2016,eta_tff_a2_2014,eta_tff_a2_2011,NA60_2009}.
 The uncertainty in the $\Lambda^{-2}_{\eta}$ value obtained
 in the present work is smaller than those of previous measurements by
 the A2 collaboration~\cite{eta_tff_a2_2014,eta_tff_a2_2011} and the NA60
 collaboration in peripheral In--In data~\cite{NA60_2009},
 but is larger than in the latest
 NA60 result, $\Lambda^{-2}_{\eta}=(1.934\pm 0.084_{\mathrm{tot}})$~GeV$^{-2}$,
 obtained from p--A collisions~\cite{NA60_2016}.

 Most of the theoretical calculations shown in
 Fig.~\ref{fig:eta_tff_cth_a2_ebst_syst_exp_th2_2x1} have already been discussed
 in Ref.~\cite{eta_tff_a2_2014}.
 The calculation by Terschl\"usen and Leupold (TL) combines the vector-meson
 Lagrangian proposed in Ref.~\cite{Lut08} and recently extended in Ref.~\cite{Ter12},
 with the Wess-Zumino-Witten contact interaction~\cite{Ter10}.
 As seen, the TL calculation lies slightly lower than the pole-approximation fit
 to the present data points, but is still in good agreement with the data points
 within their error bars.
 The calculations by the J\"ulich group, in which
 the radiative decay $\eta \to \pi^+ \pi^- \gamma$~\cite{Stollenwerk_2012}
 is connected to the isovector contributions of the $\eta \to \gamma \gamma^*$ TFF
 in a model-independent way, by using dispersion theory,
 are shown for the latest solution~\cite{Xiao_2015}, including the $a_2$-meson
 contribution in the analysis. As seen, this solution is very close
 to the present pole-approximation fit.
 The calculations by the Mainz group, which are based on a model-independent method
 using the Pad\'e approximants (initially developed for the $\pi^0$ TFF~\cite{Mas12}),
 are shown for both their previous~\cite{Esc13} and latest~\cite{Escribano_2015} solutions.
 As seen, both the solutions are very close to the present pole-approximation fit.
 However, the latest solution, also involving the previous A2 data on the
 $\eta$ TFF~\cite{eta_tff_a2_2014}, has a much smaller uncertainty.
 It is expected that adding the $|F_{\eta}(m_{e^+e^-})|^2$ results from this work
 into the corresponding calculation by the Mainz group will allow an even smaller
 uncertainty in the value for the slope parameter of the $\eta$ TFF to be obtained.

\subsection{comparison of $\omega$ results with other data and calculations}
\label{subsec:Results-III}
\begin{figure*}
\includegraphics[width=0.925\textwidth]{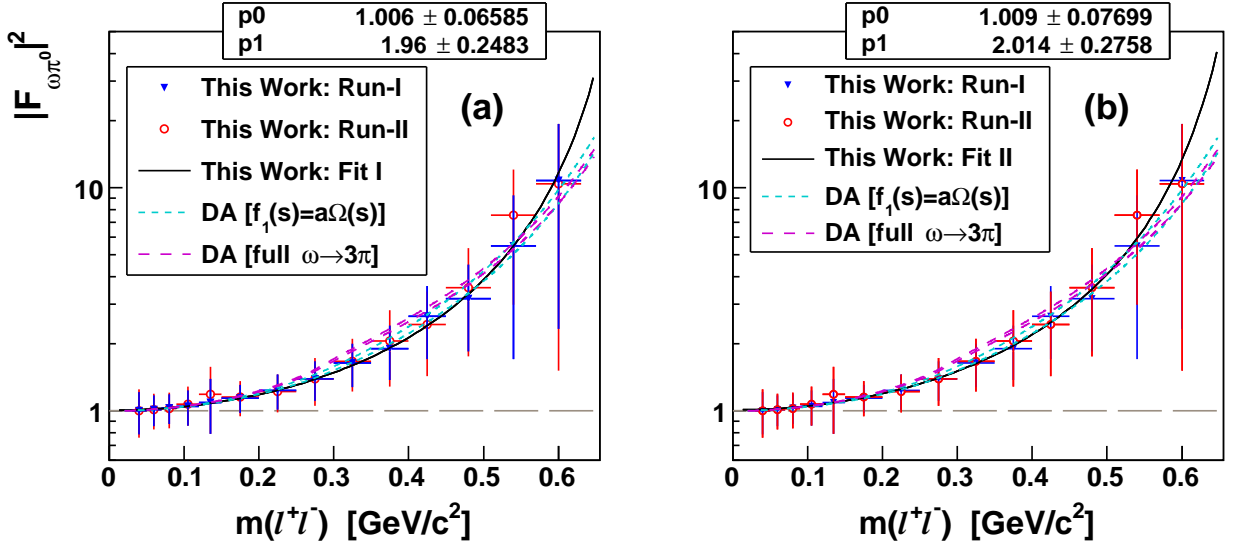}
\caption{ (Color online)
  Comparison of the $|F_{\omega\pi^0}(m_{\ell^+\ell^-})|^2$ results
 obtained individually from the analyses of Run-I (blue filled triangles)
 and Run-II  (red open circles) with each other and with the two
 solutions for the DA calculations by the Bonn
 group~\protect\cite{Schneider_2012} shown by error-band borders.
 The solution using a simplified, VMD-inspired $\omega\to 3\pi$
 partial wave $f_1(s)=\alpha\Omega(s)$ inside the dispersion integral
 is shown by cyan dashed lines, and the solution using the full rescattering
 of $3\pi$ by magenta dashed lines.
 The pole-approximation fits (black solid lines) to the results
 of Run-I and Run-II are depicted in panels (a) and (b), respectively.
 The fit parameter $p0$ reflects the general normalization of the data points,
 and $p1$ is the slope parameter $\Lambda^{-2}$.
 For a better comparison of the magnitudes of total uncertainties from the
 two data sets, the error bars of Run-I are plotted in (a) on the top of
 the error bars of Run-II, and the other way around in (b).
}
 \label{fig:omega_tff_eegg_pi0ee_cth_a2_07_vs_09_ebst_syst_thr2_2x1} 
\end{figure*}

 The individual $|F_{\omega\pi^0}(m_{e^+e^-})|^2$ results from Run-I and Run-II
 are compared in Fig.~\ref{fig:omega_tff_eegg_pi0ee_cth_a2_07_vs_09_ebst_syst_thr2_2x1}.
 Similarly to the comparison of the two individual sets of
 $|F_{\eta}(m_{\ell^+\ell^-})|^2$ results and their uncertainties,
 these experimental results are also plotted twice.
 The two-parameter fits of the individual $|F_{\omega\pi^0}|^2$ results with
 Eq.~(\protect\ref{eqn:Fm}) are shown in
 Figs.~\ref{fig:omega_tff_eegg_pi0ee_cth_a2_07_vs_09_ebst_syst_thr2_2x1}(a) and~(b),
 respectively for Run-I and Run-II.
 As seen in Fig.~\ref{fig:omega_tff_eegg_pi0ee_cth_a2_07_vs_09_ebst_syst_thr2_2x1},
 the experimental statistics for $\omega \to \pi^0e^+e^-$ decays
 in Run-I and Run-II and the level of background resulted in quite
 large total uncertainties in those $|F_{\omega\pi^0}|^2$ results,
 especially at large $m(e^+e^-)$ masses.
 Within those uncertainties, the $|F_{\omega\pi^0}|^2$ results from both
 data sets are in good agreement with each other. The same holds for
 the fit results for the normalization parameter $p0$ and
 the parameter $p1$, corresponding to $\Lambda^{-2}_{\omega\pi^0}$.  
 Despite large uncertainties in $p1=(1.96\pm0.25_{\mathrm{tot}})$~GeV$^{-2}$
 obtained for Run-I and in $p1=(2.01\pm0.28_{\mathrm{tot}})$~GeV$^{-2}$ for Run-II,
 both results indicate a lower value for $\Lambda^{-2}_{\omega\pi^0}$
 than those reported previously by Lepton-G~\cite{Lepton_G_omega}
 and NA60~\cite{NA60_2016,NA60_2009}.
 At the same time, the comparison of the individual $|F_{\omega\pi^0}(m_{e^+e^-})|^2$ results
 and their pole-approximation fits, for example, with the two different
 solutions from the dispersive analysis by the Bonn
 group~\cite{Schneider_2012} indicates no contradiction with these calculations.

\begin{figure*}
\includegraphics[width=1.\textwidth]{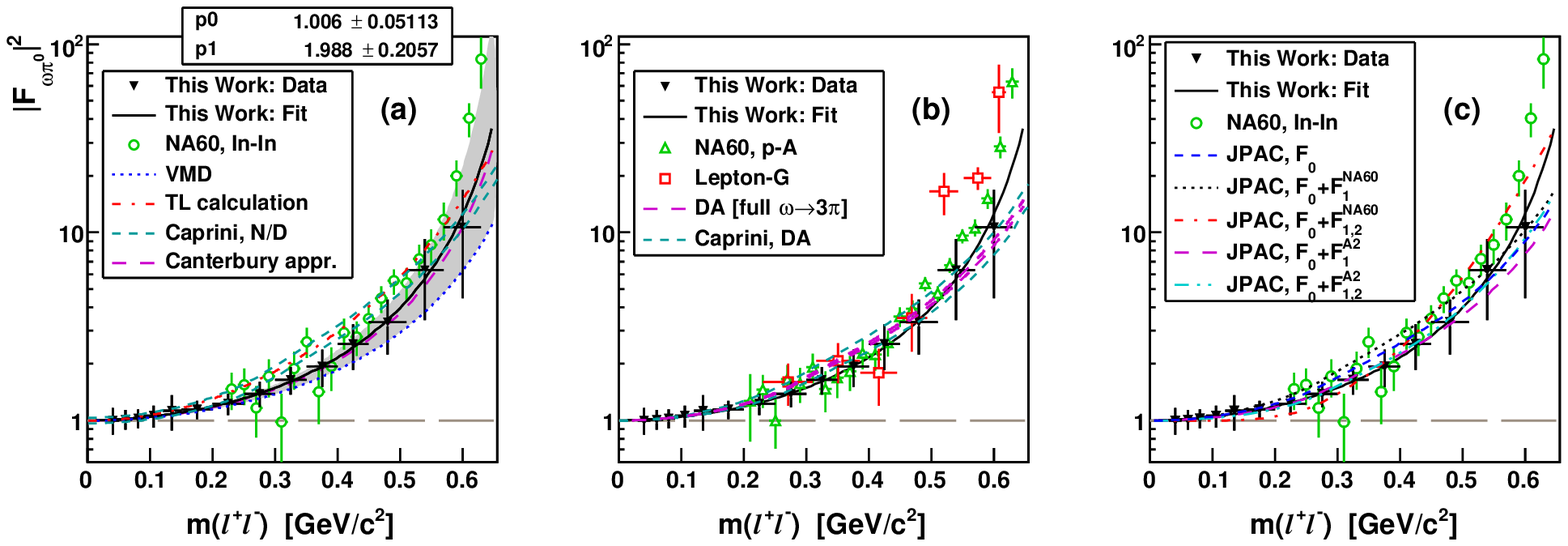}
\caption{ (Color online)
  $|F_{\omega\pi^0}(m_{\ell^+\ell^-})|^2$ results (black filled triangles) combined
 from Run-I and Run-II and their pole-approximation fit (black solid line,
 with $p0$ and $p1$ being the normalization
 and the slope parameter $\Lambda^{-2}$, respectively)
  are compared to previous measurements and various theoretical calculations.
  The results by Lepton-G~\cite{Lepton_G_omega} are shown by
  open red squares in panel (b).
  The results of NA60 obtained in peripheral In--In data~\protect\cite{NA60_2009}
  are shown by open green circles in (a) and (c),
  and from p--A collisions~\protect\cite{NA60_2016} by open green triangles in (b).  
  The VMD prediction is shown by a blue dashed line in (a).
  The calculation from Refs.~\protect\cite{TL10,Ter12} is shown
  by a red dash-dotted line in (a). 
 The DA calculation by the Bonn group~\protect\cite{Schneider_2012}
 for the full $3\pi$ rescattering
 is shown by error-band borders (magenta dashed lines) in (b).
 Upper and lower bounds by Caprini~\protect\cite{Caprini_2015} are
 shown by cyan dashed lines for two cases of the 
 discontinuity calculated with the partial-wave amplitude $f_1(t)$
 based on the improved N/D model~\cite{Koepp_1974} (a), and
 taken from Ref.~\cite{Schneider_2012} (b).
 The calculation based on a model-independent method using Canterbury
 approximants~\cite{Pere_2016} is shown by a magenta long-dashed line with
 a gray error band.
 The basic calculation (blue dashed line) from JPAC~\protect\cite{Danilkin_2015}
 and the effect from including higher order terms of the inelastic
 contributions in the $\omega\pi^0$ TFF by fitting them
 to the NA60 In--In data is shown in (c) for the solutions
 with adding one (black dotted line) and two (red dash-dotted line)
 terms. A similar effect from including higher order terms
 by fitting them to the present $|F_{\omega\pi^0}(m_{\ell^+\ell^-})|^2$ results
 is shown in (c) for the solutions with one (magenta long-dashed line)
 and with two (cyan dash-double-dotted line) terms.
}
 \label{fig:omega_tff_eegg_pi0ee_cth_a2_ebst_syst_exp_thr_3x1} 
\end{figure*}

 The $|F_{\omega\pi^0}(m_{e^+e^-})|^2$ results combined from Run-I and Run-II
 are compared to previous measurements and various theoretical calculations
 in Fig.~\ref{fig:omega_tff_eegg_pi0ee_cth_a2_ebst_syst_exp_thr_3x1}.
 The numerical values for the combined $|F_{\omega\pi^0}(m_{e^+e^-})|^2$ results
 are listed in Table~\ref{tab:omegatff}.
\begin{table*}
\caption
[tab:omegatff]{
 Results of this work for the $\omega\pi^0$ TFF, $|F_{\omega\pi^0}|^2$, as a function of
 the invariant mass $m(e^+e^-)$.
 } \label{tab:omegatff}
\begin{ruledtabular}
\begin{tabular}{|c|c|c|c|c|c|c|c|} 
\hline
 $m(e^+e^-)$~[MeV/$c^2$]
 & $40\pm10$ & $60\pm10$ & $80\pm10$ & $105\pm15$ & $135\pm15$ & $175\pm25$ & $225\pm25$ \\
\hline
 $|F_{\omega\pi^0}|^2$ 
 & $1.002\pm0.162$ & $1.011\pm0.120$ & $1.027\pm0.121$ & $1.058\pm0.140$
 & $1.126\pm0.239$ & $1.146\pm0.128$ & $1.227\pm0.161$ \\
\hline
\hline
 $m(e^+e^-)$~[MeV/$c^2$]
 & $275\pm25$ & $325\pm25$ & $375\pm25$ & $425\pm25$ & $480\pm30$ & $540\pm30$ & $600\pm30$ \\
\hline
 $|F_{\omega\pi^0}|^2$ 
 & $1.390\pm0.215$ & $1.648\pm0.279$ & $1.946\pm0.431$ & $2.553\pm0.692$
 & $3.32\pm1.08$ & $6.32\pm2.90$ & $10.63\pm6.14$ \\
\hline
\end{tabular}
\end{ruledtabular}
\end{table*}
 As seen in Fig.~\ref{fig:omega_tff_eegg_pi0ee_cth_a2_ebst_syst_exp_thr_3x1},
 the present $|F_{\omega\pi^0}(m_{e^+e^-})|^2$ results are in general agreement, within
 the error bars, with the previous measurements based on
 $\omega\to\pi^0\mu^+\mu^-$ decays. The only deviation observed is for
 the data points at the largest $m(e^+e^-)$ masses.
 The pole-approximation fit to the present $|F_{\omega\pi^0}|^2$ data points gives
\begin{equation}
 \Lambda^{-2}_{\omega\pi^0}=(1.99\pm 0.21_{\mathrm{tot}}) ~\mathrm{GeV}^{-2},
\label{eqn:Lam2_omega_this_work}
\end{equation}
 which is somewhat lower than the corresponding value obtained
 from the Lepton-G and NA60 data~\cite{Lepton_G_omega,NA60_2016,NA60_2009},
 but does not contradict them within the uncertainties.
 The uncertainty in the $\Lambda^{-2}_{\omega\pi^0}$ value obtained
 in the present work is similar to that of Lepton-G, but is
 significantly larger than the accuracy achieved by NA60.
 Meanwhile, the advantage in measuring the $\omega \to \pi^0e^+e^-$ decay is
 that the control of the overall normalization of the $|F_{\omega\pi^0}|^2$
 results is much more stringent than in the case of the $\omega\to\pi^0\mu^+\mu^-$ decay,
 which does not enable measurement at low $m(\ell^+\ell^-)$.
 The magnitude of the parameter $p0$,
 obtained from the fit to the present $|F_{\omega\pi^0}|^2$ results, indicates
 small values of systematic uncertainties due to the normalization,
 which depends on the correctness in the reconstruction of both
 the $\omega \to \pi^0e^+e^-$ and $\omega \to \pi^0\gamma$ decays
 as well as on radiative corrections for the QED differential
 decay rate at low $q$. As noted previously, the magnitude of those corrections
 is expected to be $\sim$1\%. 
 
 The basic ideas of the theoretical calculations shown in
 Fig.~\ref{fig:omega_tff_eegg_pi0ee_cth_a2_ebst_syst_exp_thr_3x1} have
 already been discussed in the Introduction.
 The calculation from Refs.~\cite{TL10,Ter12} is shown
 by a red dash-dotted line in
 Fig.~\ref{fig:omega_tff_eegg_pi0ee_cth_a2_ebst_syst_exp_thr_3x1}(a). 
 The DA calculation by the Bonn group~\cite{Schneider_2012}
 is shown in Fig.~\ref{fig:omega_tff_eegg_pi0ee_cth_a2_ebst_syst_exp_thr_3x1}(b)
 by error-band borders (magenta dashed lines) for the solution with the
 full $3\pi$ rescattering.
 The calculations by Caprini~\cite{Caprini_2015} are shown for two cases.
 Upper and lower bounds calculated with the discontinuity using
 the partial-wave amplitude $f_1(t)$ from Ref.~\cite{Schneider_2012}
 are shown in Fig.~\ref{fig:omega_tff_eegg_pi0ee_cth_a2_ebst_syst_exp_thr_3x1}(b).
 And bounds obtained with the improved N/D model~\cite{Koepp_1974} for $f_1(t)$
 are shown in Fig.~\ref{fig:omega_tff_eegg_pi0ee_cth_a2_ebst_syst_exp_thr_3x1}(a).

 There is another $\omega\pi^0$-TFF prediction translated to a simple monopole form of
 Eq.~(\ref{eqn:Fm}) with the parameter $\Lambda=(0.72\pm0.05)$~GeV,
 or $\Lambda^{-2}=(1.93\pm 0.26)$~GeV$^{-2}$~\cite{Pere_2016},
 which is depicted by a magenta long-dashed line with a gray error band in
 Fig.~\ref{fig:omega_tff_eegg_pi0ee_cth_a2_ebst_syst_exp_thr_3x1}(a).
 This calculation is based on a model-independent method using Canterbury
 approximants, which are an extension
 of the Pad\'e theory for bivariate functions~\cite{Pere_2015}.
 The parameter $\Lambda$ is obtained by requiring that the slope of the $\omega\pi^0$
 TFF in the variable $q^2$ should be the same as for the $\pi^0$ TFF,
 taking into account isospin breaking. In the approach used, the $\omega\pi^0$ TFF
 is considered as the $\pi^0$ TFF of double virtuality, with the virtuality
 of one of the photons fixed to the $\omega$-meson mass, and the other photon
 to the invariant mass of the lepton pair. The relatively large uncertainty
 in this prediction at higher $m(\ell^+\ell^-)$ is determined by the uncertainty
 in the $\pi^0$ TFF extrapolated in the region of larger $q^2$. 

 Among the calculations depicted in
 Figs.~\ref{fig:omega_tff_eegg_pi0ee_cth_a2_ebst_syst_exp_thr_3x1}(a) and (b),
 those by the Bonn group and by Caprini with the $f_1(t)$ amplitude
 from the same work~\cite{Schneider_2012} seem to be in reasonable agreement with
 the present data points. The prediction based on the method using Canterbury
 approximants is fairly close to the curve showing the data fit, but the uncertainty
 in this prediction at higher $m(\ell^+\ell^-)$ is larger, compared to the other
 calculations. Although the magnitude of the uncertainties in the present
 $|F_{\omega\pi^0}|^2$ results does not allow ruling out any of the calculations shown,
 it does challenge the understanding of the energy dependence of the $\omega\pi^0$ TFF
 at intermediate and high $q^2$.

 The calculations made for $|F_{\omega\pi^0}|^2$ by the Joint Physics Analysis Center
 (JPAC)~\cite{Danilkin_2015}
 and also the two new solutions, involving a fit to the present results,
 are shown in Fig.~\ref{fig:omega_tff_eegg_pi0ee_cth_a2_ebst_syst_exp_thr_3x1}(c).
 The basic calculation (shown by a blue dashed line) was obtained
 by using only the first term in the expansion of the inelastic
 contribution in terms of conformal variables $\omega^{i}(s)$, with its weight
 parameter determined from the experimental value for $\Gamma(\omega\to\pi^0\gamma)$.
 Other solutions were obtained by including higher-order inelastic-contribution
 terms (the next one or two orders) in the $\omega\pi^0$ TFF
 by fitting their parameters to the experimental $|F_{\omega\pi^0}|^2$ data.
 The solutions with fits to the NA60 In--In data are shown by a black dotted
 line for one additional term, and by a red dash-dotted line for two.
 The solutions with fits to the present data are shown by
 a magenta long-dashed line for one additional term, and by a cyan
 dash-double-dotted line for two. 
 As seen in Fig.~\ref{fig:omega_tff_eegg_pi0ee_cth_a2_ebst_syst_exp_thr_3x1}(c),
 the basic calculation from Ref.~\cite{Danilkin_2015} lies below
 the NA60 In--In data points at large $m(\ell^+\ell^-)$ masses, but
 comes very close to the data points of the present measurement.
 Including one more $\omega^{i}(s)$ term, with fitting its weight
 to the data, does not change much for either the NA60 In--In or
 the present results. Including two additional $\omega^{i}(s)$ terms
 in the fitting to the NA60 In--In data results in a better
 agreement with their results, but it is difficult to justify
 such a strong rise of the inelastic form factor~\cite{Caprini_2014}.
 For the present data, the solution with two additional $\omega^{i}(s)$
 terms is very close to the basic calculation, which agrees with the
 small magnitude expected for higher-order terms of the inelastic contributions. 

 Thus, the results of the present work for $|F_{\omega\pi^0}|^2$ indicate
 a better agreement with existing theoretical calculations than
 observed for previous measurements. Although the statistical
 accuracy of the present data points at large $m(\ell^+\ell^-)$
 masses does not allow a final conclusion to be drawn regarding the energy dependence
 of the $\omega\pi^0$ TFF in this region, the present $|F_{\omega\pi^0}|^2$
 results for intermediate $m(\ell^+\ell^-)$ masses obviously do not favor
 some of the calculations. 
 More measurements of the $\omega \to \pi^0e^+e^-$ decay,
 with much better statistical accuracy, especially at large $m(\ell^+\ell^-)$ masses,
 are needed to solve the problem of the inconsistency remaining between
 the calculations and the experimental data.
 Once the agreement between the theory and the experiment is established
 for the $\omega \to \pi^0\gamma^*$ TFF, or the origin of the potential
 disagreement is understood, then such data could make an
 improvement in the theoretical uncertainties, in particular the dispersive
 model-independent calculations and the Pad\'e-approximants method,
 which could then result in a better determination
 of the corresponding HLbL contribution to $(g-2)_\mu$.

 A better knowledge of radiative corrections for QED differential decay
 rates of Dalitz decays will be important for more reliable
 TFF measurements. It was checked in the present analysis
 that the correction for the QED energy dependence makes this dependence
 lower by $\sim$10\% at the largest $q$ measured.
 However, because the radiative corrections suppress the decay amplitude
 at extreme $\cos\theta^*$, taken together with the lower acceptance for those angles,
 the detection efficiency improves at large $q$. This partially compensates the impact
 from using lower QED values for measuring TFFs at large $q$.

\section{Summary and conclusions}
\label{sec:Conclusions}

 The Dalitz decays $\eta \to e^+e^-\gamma$ and $\omega\to \pi^0e^+e^-$ have
 been measured in the $\gamma p\to \eta p$ and $\gamma p\to \omega p$ reactions,
 respectively, with the A2 tagged-photon facility at the Mainz Microtron, MAMI.
 The value obtained for the slope parameter of the $\eta$ e/m TFF,
 $\Lambda^{-2}_{\eta}=(1.97\pm 0.11_{\mathrm{tot}})$~GeV$^{-2}$,
 is in good agreement with previous measurements of
 the $\eta\to e^+e^-\gamma$ and $\eta \to \mu^+\mu^-\gamma$ decays,
 and the $|F_{\eta}|^2$ results are in good agreement
 with recent theoretical calculations.
 The uncertainty obtained in the value of $\Lambda^{-2}_{\eta}$
 is lower than in previous results based on
 the $\eta\to e^+e^-\gamma$ decay and the NA60 result based on
 $\eta\to\mu^+\mu^-\gamma$ decays from peripheral In--In collisions.
 The value obtained for $\omega$,
 $\Lambda^{-2}_{\omega\pi^0}=(1.99\pm 0.21_{\mathrm{tot}})$~GeV$^{-2}$,
 is somewhat lower than previous measurements based on the
 $\omega\to\pi^0\mu^+\mu^-$ decay.
 The results of this work for $|F_{\omega\pi^0}|^2$ are in better agreement
 with theoretical calculations than the data from earlier experiments.
 However, the statistical accuracy of the present data points
 at large $m(e^+e^-)$ masses does not allow a final conclusion to be drawn
 about the energy dependence in this region.
 More measurements of the $\omega \to \pi^0e^+e^-$ decay,
 with much better statistical accuracy, especially at large $m(e^+e^-)$ masses,
 are needed to solve the problem in the inconsistency remaining between
 the calculations and the experimental data.
 Compared to the $\eta\to \mu^+\mu^-\gamma$ and $\omega\to\pi^0\mu^+\mu^-$ decays,
 measuring $\eta\to e^+e^-\gamma$ and $\omega\to \pi^0e^+e^-$ decays
 gives access to the TFF energy dependence at low momentum transfer, which is
 important for data-driven approaches calculating the corresponding rare decays
 and the HLbL contribution to $(g-2)_\mu$.

\section*{Acknowledgments}

 The authors wish to acknowledge the excellent support of the accelerator group and
 operators of MAMI.
 We would like to thank Bastian Kubis, Stefan Leupold, Pere Masjuan, and Irinel Caprini
 for useful discussions and continuous interest in the paper.
 This work was supported by the Deutsche Forschungsgemeinschaft (SFB443,
 SFB/TR16, and SFB1044), DFG-RFBR (Grant No. 09-02-91330), the European Community-Research
 Infrastructure Activity under the FP6 ``Structuring the European Research Area''
 program (Hadron Physics, Contract No. RII3-CT-2004-506078), Schweizerischer
 Nationalfonds (Contract Nos. 200020-156983, 132799, 121781, 117601, 113511),
 the U.K. Science and Technology Facilities Council (STFC 57071/1, 50727/1),
the U.S. Department of Energy (Offices of Science and Nuclear Physics,
 Award Nos. DE-FG02-99-ER41110, DE-FG02-88ER40415, DE-FG02-01-ER41194)
 and National Science Foundation (Grant Nos. PHY-1039130, IIA-1358175),
 INFN (Italy), and NSERC (Canada).
 We thank the undergraduate students of Mount Allison University
 and The George Washington University for their assistance.

\end{document}